\documentclass[10pt,letterpaper,conference]{IEEEtran}
\usepackage{amsmath,amssymb,amsfonts}
\usepackage{gensymb}
\usepackage{array}
\usepackage{algorithmic}
\usepackage{graphicx}
\usepackage{textcomp}
\usepackage{xcolor}
\usepackage{xspace}
\usepackage{enumitem}
\usepackage{url}
\usepackage{breakurl}
\usepackage{caption}
\usepackage{subcaption}
\usepackage{multirow}
\usepackage[normalem]{ulem}
\usepackage[numbers,sort&compress]{natbib}
\usepackage[linesnumbered,ruled,lined,boxed]{algorithm2e}
\graphicspath{{figures/}}
\SetKwInput{KwInput}{Input}                % Set the Input
\SetKwInput{KwOutput}{Output}              % set the Output

\newcolumntype{L}[1]{>{\raggedright\let\newline\\\arraybackslash\hspace{0pt}}m{#1}}
\newcolumntype{C}[1]{>{\centering\let\newline\\\arraybackslash\hspace{0pt}}m{#1}}
\newcolumntype{R}[1]{>{\raggedleft\let\newline\\\arraybackslash\hspace{0pt}}m{#1}}

\newcommand{\coloroverride}[2]{\textcolor{#1}{#2}}
\renewcommand{\coloroverride}[2]{\textcolor{black}{#2}}

\newcommand{\done}[1]{\coloroverride{blue}{#1}} %Blue to review
\newcommand{\chk}[1]{\coloroverride{red}{#1}} %Red to check

\newcommand{\ie}{\textit{i.e.,}\xspace}

\usepackage{csquotes}
\usepackage{longtable}
\usepackage{graphicx}   
\usepackage[caption=false,listofformat=subsimple,labelformat=simple]{subfig}

\begin{document}

\title{Evaluation of Indoor/Outdoor Sharing in the Unlicensed 6 GHz Band}

% IEEE Author block
\author{ 
\IEEEauthorblockN{
Seda Dogan-Tusha\IEEEauthorrefmark{1},
Armed Tusha\IEEEauthorrefmark{1},
Muhammad Iqbal Rochman\IEEEauthorrefmark{1},
Hossein Nasiri\IEEEauthorrefmark{1},
Joshua Roy Palathinkal\IEEEauthorrefmark{1},\\
Mike Atkins \IEEEauthorrefmark{2},
and Monisha Ghosh\IEEEauthorrefmark{1}\\
\IEEEauthorblockA{
\IEEEauthorrefmark{1}Department of Electrical and Electronics Engineering, University of Notre Dame, South Bend, IN, USA\\
\IEEEauthorrefmark{2}Office of Information Technology, University of Notre Dame, South Bend, IN, USA}
\IEEEauthorblockA{
Email: \{stusha, atusha, mrochman, hnasiri2, jpalathi, matkins, mghosh3\}@nd.edu}}
}

% Anonymous author for double-blind review
% \author{\IEEEauthorblockN{Submission \#XXXXX}}

\maketitle

%%%%%%%%%%%%%%%%%%%%%%%%%%%%%%%%%%%%%%%%%%%%%%%%%%

\begin{abstract}
% This paper presents a comprehensive characterization of 6 GHz Wi-Fi performance in both indoor and outdoor environments, using measurements data collected at a dense 6 GHz Wi-Fi deployment in the Notre Dame stadium. We investigate the mutual impact of indoor and outdoor deployments, analyzing potential interference and coexistence challenges.  Additionally, we examine the utilization of the 6 GHz band compared to the established 5 GHz band, assessing key metrics such as the percentage of connections and achieved throughput. Our findings reveal that deployments of 6 GHz standard power outdoors may lead to interference to low-power indoor deployments, depending on the building material. This underscores the importance of indoor-outdoor separation for optimal 6 GHz Wi-Fi performance and future hybrid indoor-outdoor sharing schemes.

Standard Power (SP) Wi-Fi 6E in the U.S. is just beginning to be deployed outdoors in the shared but unlicensed 6 GHz band under the control of an Automated Frequency Coordination (AFC) system to protect incumbents, while low-power-indoor (LPI) usage has been steadily increasing over the past 2 years. In this paper, we present the first comprehensive measurements and analyses of a SP Wi-Fi 6E deployment at the University of Notre Dame's football stadium, with 902 access points and a seating capacity of 80,000, coexisting with LPI deployments in adjacent buildings. Measurement campaigns were conducted during and after games, outdoors and indoors to fully characterize the performance of SP Wi-Fi 6E, interactions between SP and LPI and potential for interference to incumbents. Our main conclusions are: (i) in a very short time of about 2 months, the percentage of Wi-Fi 6E client connections is already 14\% indicating rapid adoption, (ii) dense SP operation outdoors can negatively impact LPI deployments indoors, depending on building loss, indicating the need to carefully consider hybrid indoor-outdoor sharing deployments, and (iii) spectrum analyzer results indicate an aggregate signal level increase \chk{of approximately 10 dB} in a Wi-Fi channel during peak usage which could potentially lead to interference since the AFC does not consider aggregate interference when allocating permitted power levels. These results from real-world deployments can inform spectrum policy in other bands where similar sharing mechanisms are being considered, such as 7.125 - 8.4 GHz.


\end{abstract}

%%%%%%%%%%%%%%%%%%%%%%%%%%%%%%%%%%%%%%%%%% 

\section{Introduction}\label{introduction}

\begin{figure}
    \centering
    \includegraphics[width=\linewidth]{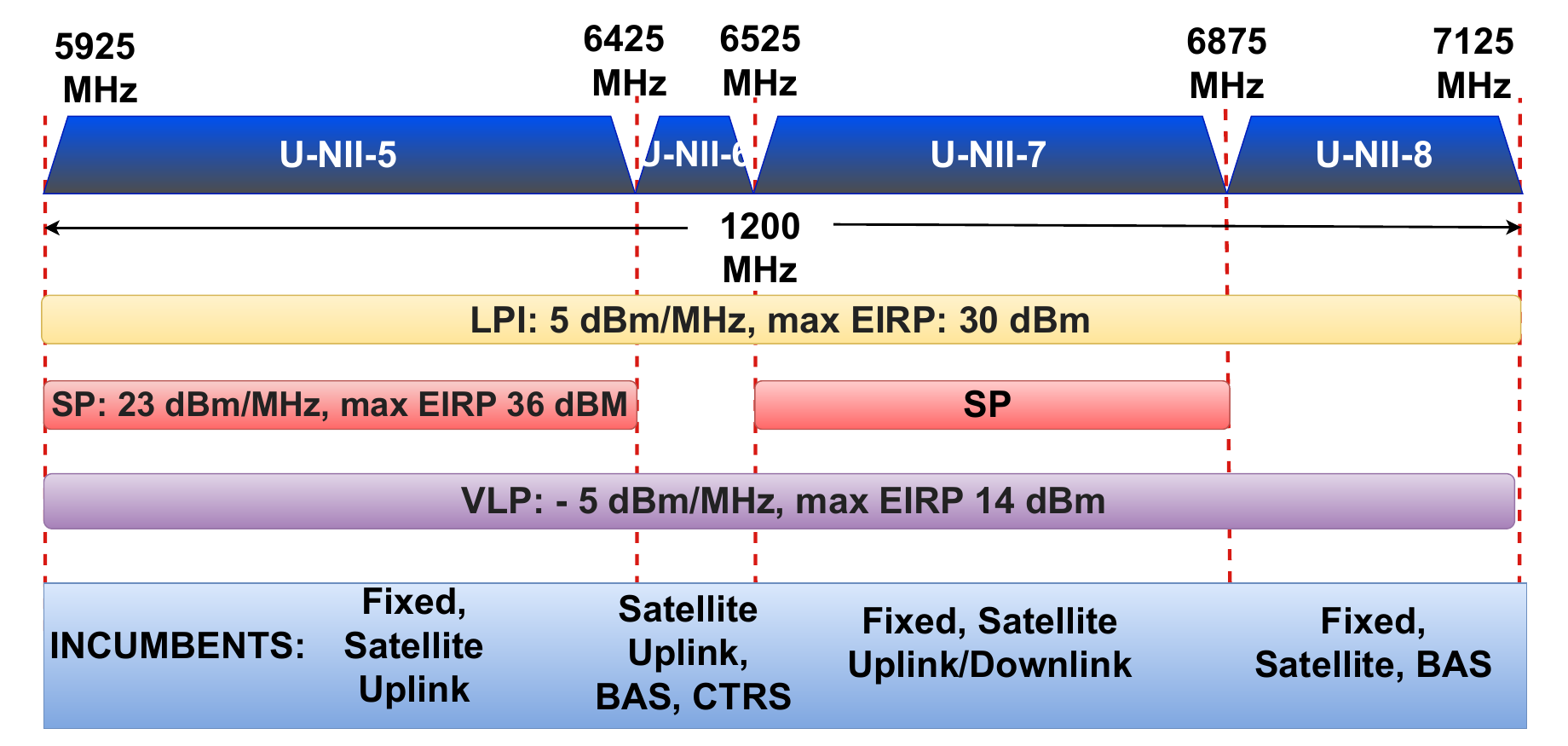}
    \caption{Spectrum Bands in 6 GHz.\color{black}}
    \label{fig:6ghz_freq_chart}
    \vspace{-1em}
\end{figure}

Since 2020, the 6 GHz band (5.925 - 7.125 GHz) has been adopted either partially or fully for unlicensed but shared use in many parts of the world~\cite{WiFi2}. Since incumbent use and regulations differ across these regions, our discussion in this paper will be restricted to the U.S. where the entire 1.2 GHz has been allocated for unlicensed use, and is composed of four distinct Unlicensed National Information and Infrastructure (U-NII) bands as shown in Fig.~\ref{fig:6ghz_freq_chart}: U-NII-5 (5.925 - 6.425 GHz), U-NII-6 (6.425 - 6.525 GHz), U-NII-7 (6.525 - 6.875 GHz) and U-NII-8 (6.875 - 7.125 GHz). These demarcations are based on the predominant incumbents in each U-NII-band: fixed links used by entities such as public utilities are deployed in U-NII-5 and U-NII-7 while U-NII-6 and U-NII-8 are used by mobile television relays and electronic news gathering services. In order to protect these incumbents from harmful interference, while maximizing utilization, the U.S. has adopted three power regimes for operation in the band: (i) Standard Power (SP), indoors or outdoors, under the control of an Automated Frequency Coordination (AFC) system, with power spectral density (PSD) of 23 dBm/MHz and 36 dB maximum Effective Isotropic Radiated Power (EIRP), (ii) Low Power Indoor (LPI), indoors only, with PSD of 5 dBm/MHz and maximum EIRP of 30 dBm, without AFC, and (iii) Very Low Power (VLP), indoors or outdoors, without AFC, with PSD of -5 dBm/MHz and maximum EIRP of 14 dBm \cite{FCC2,FCC5}. SP usage is permitted only in U-NII-5 and U-NII-7 since fixed link incumbents can be protected by an AFC while LPI and VLP are permitted in U-NII-5 to U-NII-8. Thus, the rules enable a wide variety of applications: wide-area coverage using SP with AFC, local indoor use with LPI and personal/portable use such as smart glasses using VLP. However, there is no real-world evaluation of how these different devices, with different power levels and operating regimes coexist with each other. In this paper, we provide detailed measurements and analyses of coexistence of a dense outdoor SP deployment with a dense, indoor LPI deployment where both use the latest Wi-Fi 6E devices: we believe that this is the first such analysis of a real-world deployment.

Such studies can inform ongoing discussions about sharing in the 6 GHz band as well as in the 7.125 - 8.4 GHz band. Ofcom in the U.K. is evaluating hybrid sharing in the upper 6 GHz band, with outdoor cellular and indoor Wi-Fi coexisting in the same frequency band~\cite{ofcom20236ghz}. The National Telecommunications and Information Administration (NTIA) in the U.S. is actively exploring shared spectrum access in the mid-band frequency ranges 3.1 - 3.45 GHz and 7.125 - 8.4 GHz~\cite{ntia2023strategy}. One of the primary incumbents in the 7.125 - 8.4 GHz band is federal fixed links, similar to the commercial ones deployed in the 6 GHz band, making the study of real-world deployed systems in 6 GHz particularly useful.

% This growing interest in location-dependent sharing highlights the importance of understanding coexistence between SP and LPI deployments \cite{dogan2023evaluating}.

Since LPI devices do not require an AFC system for operation, deployments have been steadily increasing since 2022~\cite{dogan2023evaluating,dogan2023indoor}. However, SP deployments could only begin once the AFC systems were deployed and certified by the Federal Communications Commission (FCC) in February 2024~\cite{FCC4}. Similar to the \done{Spectrum Access System} (SAS) used in Citizens Band Radio Service (CBRS) to protect incumbent use, the 6 GHz AFC is a cloud-based system that reports the available channels and allowed transmit power based on the location of the unlicensed device, while ensuring that interference to any incumbent in the area is limited to an interference-to-noise (I/N) ratio of -6 dB. The AFC extracts parameters of fixed links in U-NII-5 and U-NII-7 through the FCC's Universal Licensing System (ULS)~\cite{FCC_ULS} and determines the I/N based on a prescribed propagation model. For instance, Cambium's 6 GHz Spectrum Availability Tool allows network operators to pre-check the available channels at any location that they choose \cite{cambium}. Each SP AP in a Wi-Fi 6E network must directly connect to the AFC system for coordinating and ensuring safe spectrum sharing with the incumbents in the band. 
\chk{In the 6 GHz band, AFC protects incumbents by ensuring that unlicensed devices, such as Wi-Fi 6E and future Wi-Fi 7 systems, do not interfere with licensed users like fixed microwave links and satellite services. AFC systems  adjust the frequency and power levels of unlicensed devices to prevent harmful interference to the incumbents.}

\begin{figure}
    \centering  \includegraphics[width=\linewidth, height=0.55\linewidth]{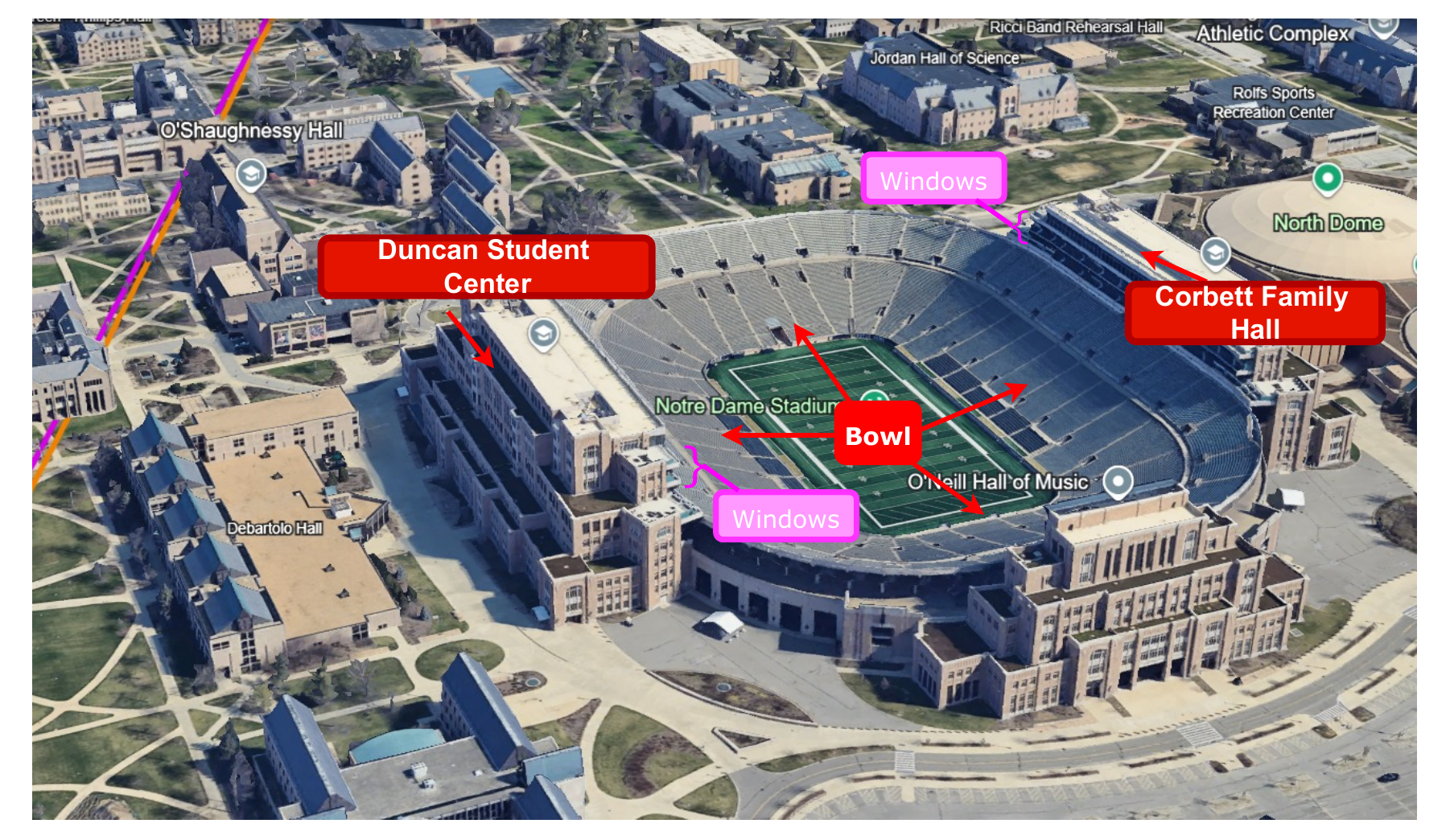}
    \caption{Measurement locations.}
    \label{fig:NDSmeasurment}
    \vspace{-1em}
\end{figure}

In this paper, we leverage a recently deployed dense, outdoor SP Wi-Fi 6E infrastructure in the football stadium at the University of Notre Dame (ND)~\cite{ndnews}, along with indoor LPI Wi-Fi 6E in two buildings attached to the stadium to perform extensive measurements and detailed analyses of coexistence between the two deployments as well as evaluate the probability of interference to incumbents due to aggregate interference from the dense SP and LPI deployment. Our main contributions are:

\noindent $\bullet$
A first of its kind detailed data set of indoor and outdoor measurements of SP and LPI APs.

\noindent $\bullet$
Comparison of 6 GHz usage and 5 GHz usage when the stadium is at full capacity with 80,000 attendees.

\noindent $\bullet$
Detailed analyses of coexistence of outdoor SP with indoor LPI under different conditions: fully occupied stadium and empty stadium.

\noindent $\bullet$
Spectrum analyzer measurements quantifying the increase in aggregate interference when the stadium is at full capacity.

\begin{figure}
    \begin{subfigure}{.48\linewidth}
    \includegraphics[width=\linewidth]{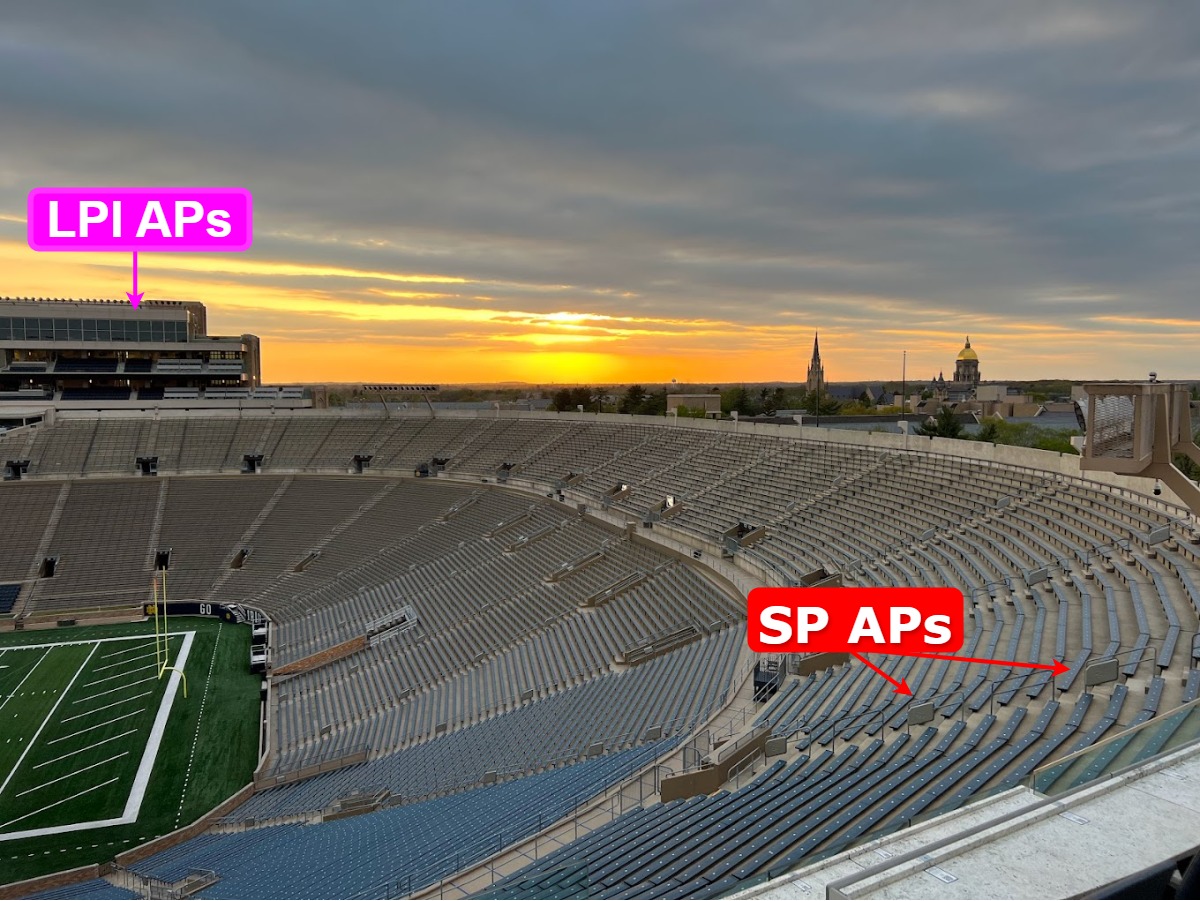}
    \caption{Stadium bowl.}
    \label{fig:bowlST}
    \end{subfigure}
    \hfill
    \begin{subfigure}{.48\linewidth}
\includegraphics[width=\linewidth]{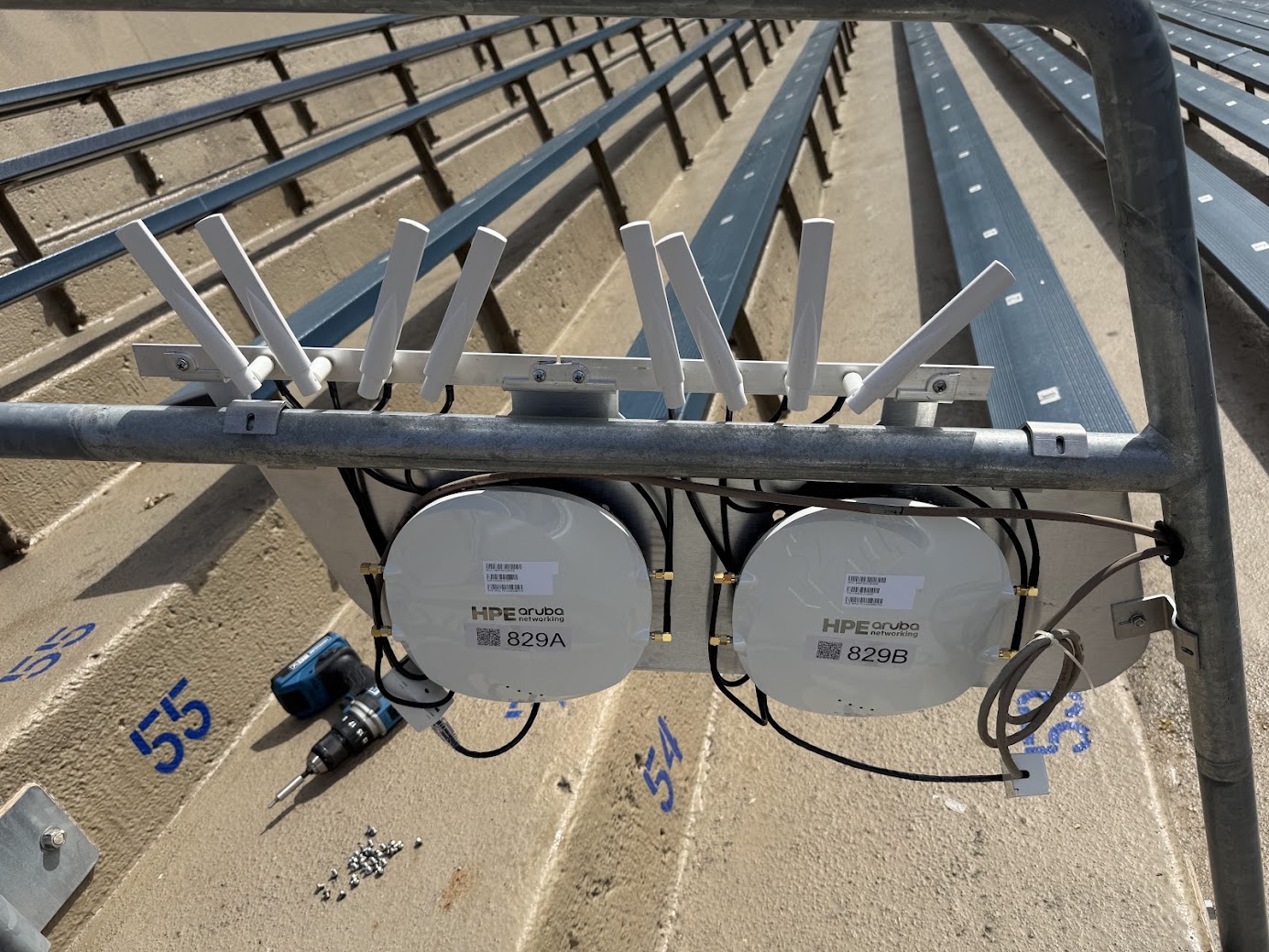}
    \caption{APs mounted on a handrail.}
    \label{fig:APfig}
    \end{subfigure}
    % \vspace{-.5em}
    \caption{Wi-Fi 6E deployment at Notre Dame stadium bowl.}
    \label{fig:bowldep}
 \end{figure}

%%%%%%%%%%%%%%%%%%%%%%%%%%%%%%%%%%%%%%%%%% 
\section{Deployment, Tools, and Methodology}

Extensive measurement campaigns were conducted in the ND stadium and two adjoining buildings, Duncan Student Center and Corbett Family Hall as shown in Fig.~\ref{fig:NDSmeasurment}, using specific tools and methodologies which will be described in detail in this section.

\begin{figure}
    \centering
    \includegraphics[width=\linewidth]{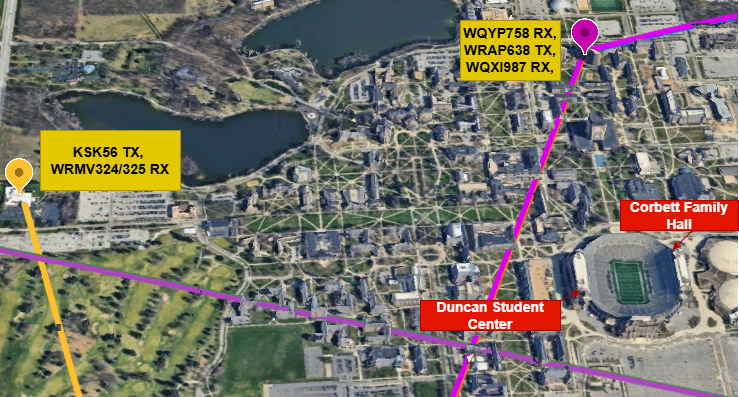}
   % {figures/und_stadium.png}
    \caption{\color{black}Fixed links.\color{black}}
    \label{fig:fixedlinks}
    \vspace{-1em}
\end{figure}

%%%%%%%%%%%%%%%%%%%%%%%%%%%%%%%%%%%%%%%%%%%%%%%%%%%%%%%%%%%%%%%%%%

\begin{table}[t]
	\caption{Measurement tools and devices.}
	\centering
	\small	\renewcommand{\arraystretch}{1.1}
	\begin{tabular}{|C{1.8 cm}|C{3cm}|C{2.6cm}|} 
 \hline
 \textbf{App./Tool} & \textbf{Features}  & \textbf{Devices} \\
  \hline \hline  SigCap &  BSSID, Tx power, \# of connected clients, frequency, channel utilization, RSSI, GPS, Tx and Rx link-speeds when connected & 3 $\times$ Google P8, \newline 2 $\times$ Samsung S22, \newline 1 $\times$ Samsung S24 \\
     \hline
   SA & Spectrum power captured over 100 kHz bins & 1 $\times$ Keysight FieldFox, Model N9951A Handheld SA\\
\hline
\end{tabular}
\label{tab:FeatureDevices} 
\vspace{-1em}
\end{table}

%
% Subsection
%
\subsection{Deployment Overview}\label{sec:deployments}
% Notre Dame stadium is the home of the University of Notre Dame's football team, the Fighting Irish, representing one of the most iconic sports architecture worldwide, shown in Fig.~\ref{fig:NDSmeasurment}. 

The ND stadium shown in Fig.~\ref{fig:NDSmeasurment} consists of an open bowl area and three adjacent buildings anchored to the south, east, and west sides: O'Neill Hall, Corbett Family Hall, and Duncan Student Center respectively. Detailed measurements were made outdoors in the bowl and indoors in Corbett and Duncan. Both buildings have nine floors, of which floors 1 - 6 are attached to the bowl with concrete walls separating them, while floors 7 - 9 rise above the bowl with large windows overlooking the stadium. Thus, there are three distinct environments: outdoors in the bowl area, indoors near windows in floors 7 - 9 of Corbett and Duncan, and indoor interior in floors 1 - 6 of the two buildings. 

Fig. \ref{fig:bowlST} shows the stadium bowl with a capacity of nearly 80k seats, which translates to roughly 1 person per square meter. About 900 SP Wi-Fi 6E APs \chk{(Aruba AP-634)} are installed outdoors in the stadium bowl: two SP APs are placed within a case and mounted on the handrail that splits the stadium sections as shown in Fig. \ref{fig:APfig}. Each AP has 4 antennas \chk{(Aruba AP-ANT-311)}, with the middle two antennas installed perpendicular to the railing and the outside antennas are about 40 degrees off. \chk{Each antenna have a peak gain of 3 dBi in 2.4 GHz, and 6 dBi in both 5 and 6 GHz.} About 400 LPI APs \chk{(Aruba AP-635)} are installed indoors at Corbett and Duncan. \chk{These LPI APs are mounted facing down in the ceiling, with integrated omnidirectional antennas. The peak gain of the antenna is 4.6 dBi in 2.4 GHz, 7.0 dBi in 5 GHz and 6.3 dBi in 6 GHz.} On game-day each SP AP only broadcasts on one SSID, ND-guest, whereas on post-game-days two SSIDs are broadcast per AP: ND-guest and eduroam. Since most spectators cannot connect to eduroam, this deployment maximizes the utilization by outside visitors to the stadium. It should also be noted that each AP broadcasts ND-guest on 5 GHz, using 20 MHz channels, and 6 GHz, using 80 MHz channels.  

\begin{figure} 
    \includegraphics[width=\linewidth]{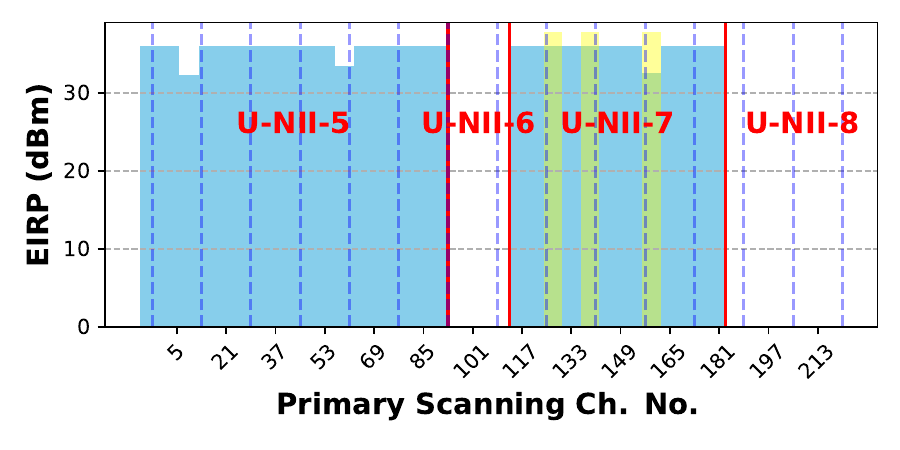} 
    \caption{AFC plot from Cambium. Yellow bars are the receive frequencies of the three fixed links in the vicinity.}
    \label{fig:eirp_fixedlink}
    \vspace{-1em}
\end{figure}

With approximately 1300 Wi-Fi 6E APs installed indoors and outdoors in a small area, aggregate interference could potentially affect incumbents. There are five fixed links in the vicinity of the stadium, as shown in Fig.~\ref{fig:fixedlinks}. Three of these fixed links are deployed on the roof of Grace Hall, a building on the ND campus. Two of these links are receivers (WQXI987 at frequency 6655 MHz and WQYP758 at 6595 MHz with 30 MHz bandwidth) with transmitters positioned at a significant distance away from the campus, and one is a transmitter, WRAP638 on 6755 MHz with the receiver in South Bend. These frequencies are shown as yellow bars in Fig.~\ref{fig:eirp_fixedlink}. None of these links have beams that are close enough to the stadium to appreciably affect the allowed power as seen in Fig.~\ref{fig:eirp_fixedlink}. However, it should be noted that the AFC does not consider aggregate interference in the calculation and, as will be seen later, we note a significant increase in signal strength when the stadium is at full capacity.

% WRAP638, at frequencies 6755 MHz and 6815 MHz, is the only transmitter fixed link on campus with the link direction passing near the Duncan Student Center building.

\subsection{Tools}
To capture detailed Wi-Fi signal information, \chk{we conducted walking measurements} using six smartphones capable of Wi-Fi 6E operation in 6~GHz : three Pixel 8, one Samsung S24, and two Samsung S22 phones as listed in Table~\ref{tab:FeatureDevices}. As we utilized varying phone models, we can expect differences in measurement sensitivity. However, we kept the same set of phones in all of our measurements to ensure consistency.

Each phone is equipped with SigCap~\cite{sigcap} which extracts various signal parameters from the modem chip through Android APIs without requiring root access. SigCap passively collects measurements every 5 seconds, with each data entry including a timestamp, GPS coordinates, and detailed information on available Wi-Fi, 4G and 5G channels. In particular, SigCap decodes Wi-Fi beacon frames transmitted by the AP to capture information such as the Service Set Identifier (SSID), Basic Service Set Identifier (BSSID), frequency, bandwidth, primary channel, transmit power, number of connected clients, Received Signal Strength Indicator (RSSI), and channel utilization. SigCap can also generate various types of traffic for active measurements such as iperf~\cite{iperf}, HTTP download, and ping, which we use to ensure full channel utilization for accurate network measurements. In total, we extracted 787,777 GPS and timestamped entries as CSV files for further analysis.
 \begin{figure*}
    \begin{subfigure}{.30\linewidth}
    \includegraphics[width=\linewidth]{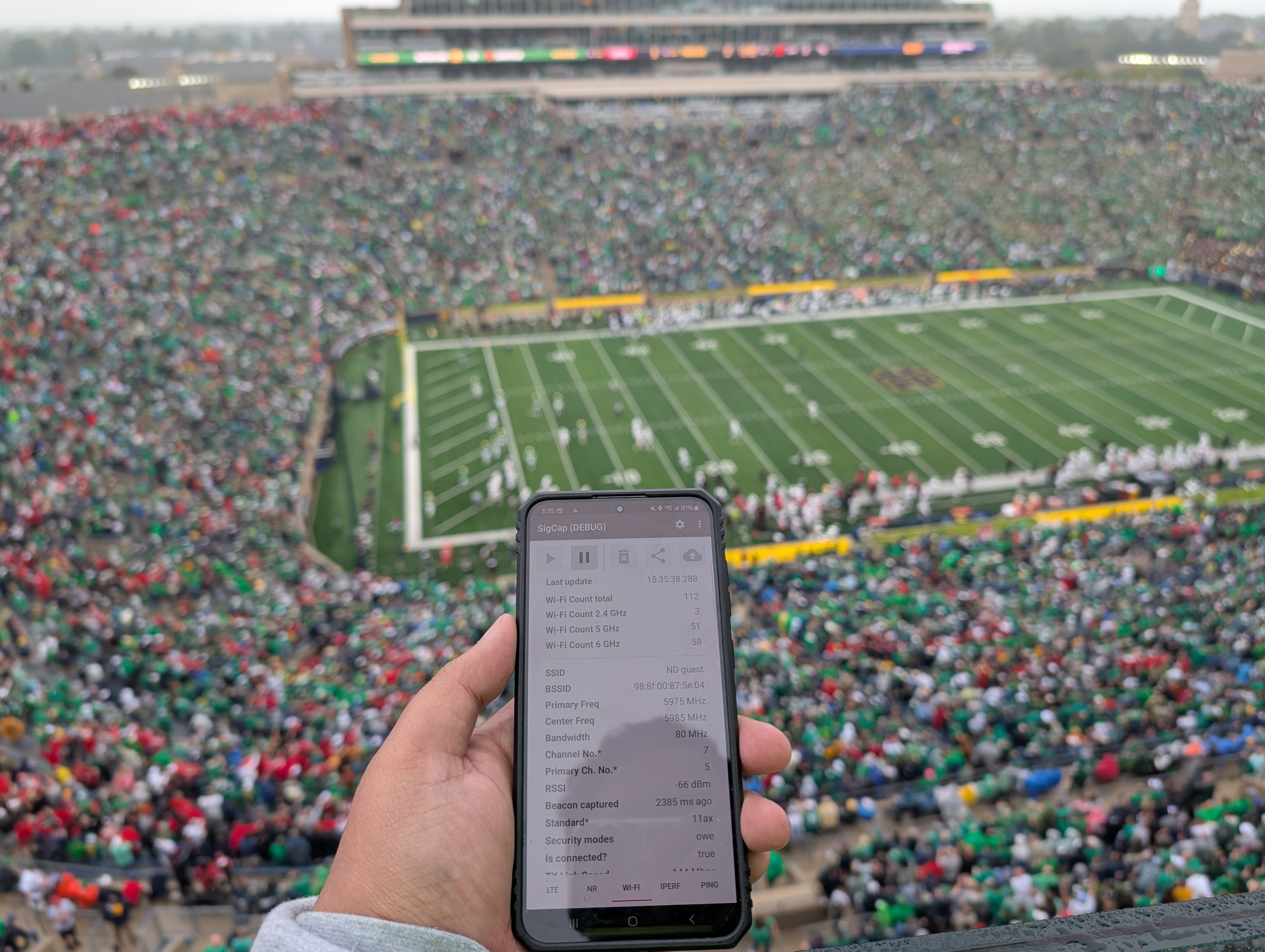}
    \caption{Outdoor walking campaign}
    \label{fig:outdoor_rsrp_ent}
    \end{subfigure}
    \hfill
    \begin{subfigure}{.30\linewidth}
    \includegraphics[width=\linewidth]{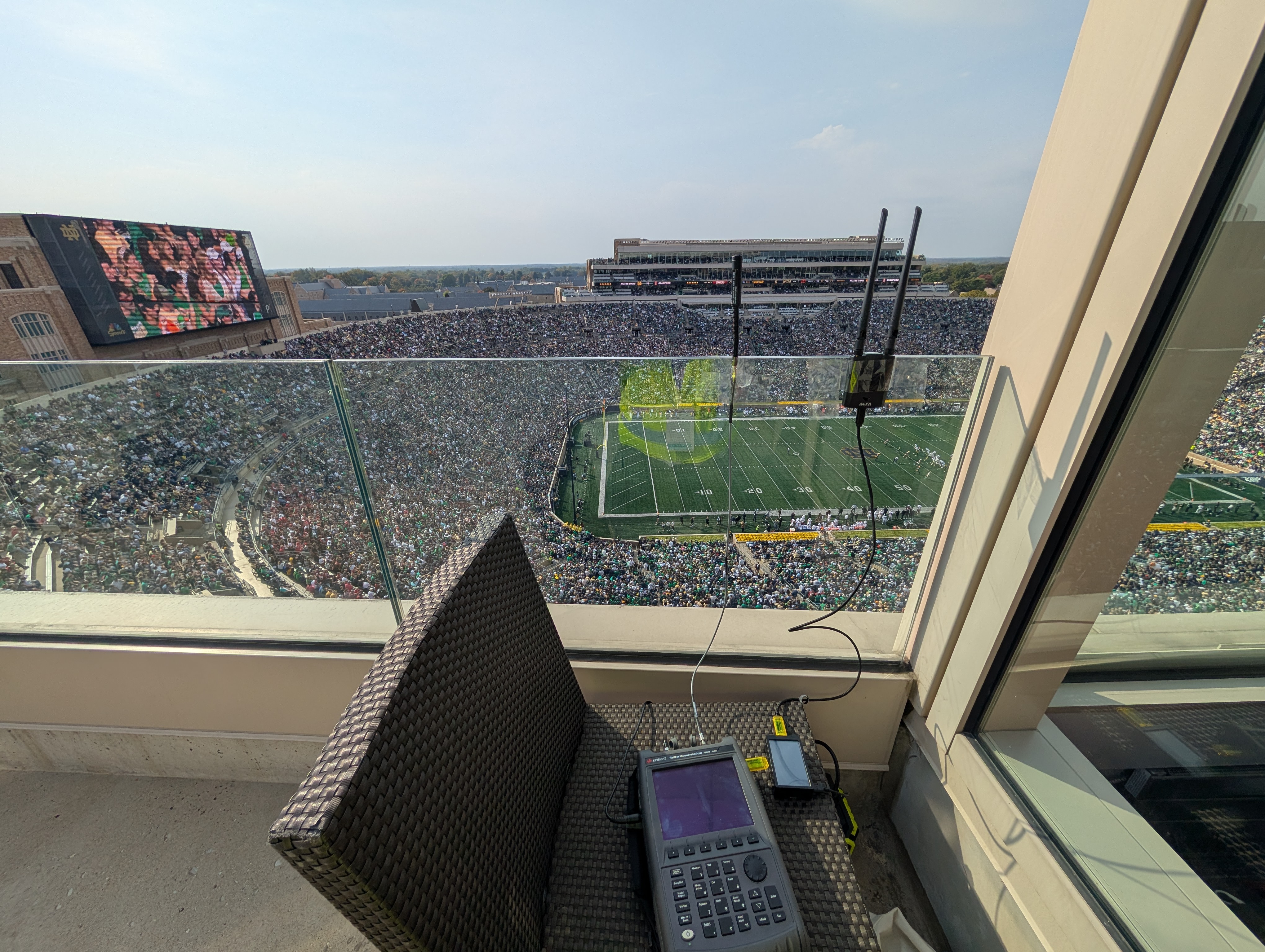}
    \caption{Outdoor fixed location campaign}
    \label{fig:outdoor_SA}
    \end{subfigure}
   \hfill
    \begin{subfigure}{.30\linewidth}
    \includegraphics[width=\linewidth]{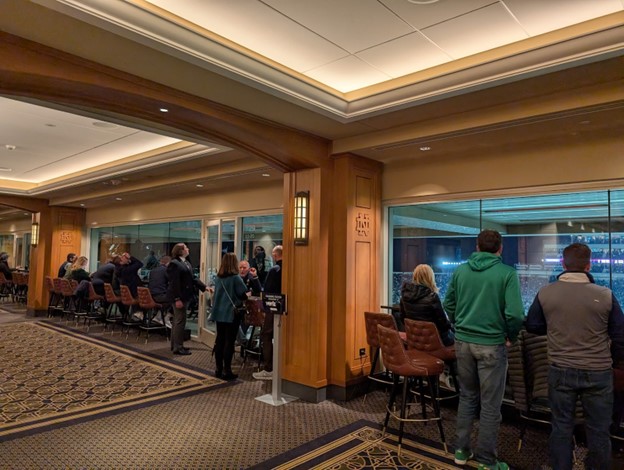}
    \caption{Indoor walking campaign}
    \label{fig:indoor_rsrp_dc}
    \end{subfigure}
    % \vspace{-.5em}
    \caption{Measurement campaigns at Notre Dame stadium.}
    \label{fig:outdoor_rsrps}
    \vspace{-1em}
 \end{figure*}

Accurate spectrum measurements were collected with a spectrum analyzer (SA), the Keysight FieldFox Handheld Real-time Spectrum Analyzer model N9951A. Due to the maximum bandwidth limitation of 100 MHz, we measured power sequentially over the fourteen 80 MHz bandwidth Wi-Fi channels. Power is measured over 100 kHz bins (resolution bandwidth): there are 821 bins per sample. The SA is set with a reference level of -24 dBm, attenuation of 20 dB, pre-amp on and IF path set to auto at 100 MHz. It provides a 100\% probability of intercept at 15.4 microseconds. Over the two days of measurements, we captured 59,066 samples with a sampling rate of 20 ms.
   
\subsection{Methodology}
Multiple measurement campaigns were undertaken from September to November 2024 to evaluate the coexistence performance of outdoors SP and indoors LPI Wi-FI 6E APs, during game-days (GD) and post-game-days (PGD).
% in an unbiased manner, as listed in Table VV. In this work, we consider only the campaigns that offer the most comprehensive coverage both indoors and outdoors; large area, more measurement points, number of SP/LPI APs seen, and other detailed observations. 
The measurements discussed in this work were carried out during the game-day on November 9, 2024, with a full stadium, and post-game-day on November 25, 2024 when the stadium was empty. Measurements were taken while walking and in a fixed location, indoors and outdoors, with the same areas covered during both measurement days. 

\textbf{Outdoor Measurements:} Game-day and post-game-day walking measurements were conducted in the bowl \chk{using phones equipped with SigCap}, as shown in Fig. \ref{fig:outdoor_rsrp_ent}. Since the stadium was at full capacity, outdoor walking measurements on game-day could only be taken along the designated walking areas between the bowl sections, up and down the stairs.

The spectrum analyzer, with an antenna designed for 4-8 GHz, was placed on the balcony of the 9th floor of Duncan Student Center, for both game-day and post-game-day measurements, as shown in Fig. \ref{fig:outdoor_SA}. The antenna was positioned to collect measurements in a stationary outdoor environment.

\textbf{Indoor Measurements:} Due to restricted access, indoor game-day and post-game-day walking measurements were conducted \chk{using SigCap phones}, in floors 7 - 9 of Corbett and floors 1, 2, 7 - 9 of Duncan. As described previously, the 1st and 2nd floors of Duncan are connected to the stadium by concrete walls, with no windows facing the stadium, floors 7 - 9 of Corbett and Duncan feature a long hallway with large double-pane low-E windows and glass doors facing the stadium bowl, as shown in Fig.~\ref{fig:indoor_rsrp_dc}. 
For the rest of the paper, floors 1 -2 of Duncan will be referred to as \enquote{indoor interior (II)}, floors 7 -9 of Duncan and Corbett will be referred to as \enquote{indoors near windows (INW)} and stadium bowl as \enquote{outdoors (Out)}, as listed in Table \ref{tab:meas_env}.

\begin{table}
\renewcommand{\arraystretch}{1.3}
\centering
\caption{Description of measurement environments.}
\begin{tabular}{|c|p{4cm}|}
\hline
\textbf{Environment} & \textbf{Description} \\ \hline \hline
\textbf{Out (Stadium Bowl)} & The open area of the stadium bowl, with SP deployments \\ \hline
\textbf{II (Indoor Interior)} & Floors 1 - 2, Duncan Student Center, with concrete walls and no windows facing the stadium. \\ \hline
\textbf{INW (Indoors near Windows)} & Floors 7 - 9, Duncan Student Center and Corbett Family Hall, featuring long hallways with large double-pane low-E windows and glass doors to the stadium bowl. \\ \hline
\end{tabular}
\label{tab:meas_env}
\vspace{-1em}
\end{table}

\begin{figure}
\begin{subfigure}{\linewidth}
    \centering   \includegraphics[width=\linewidth]{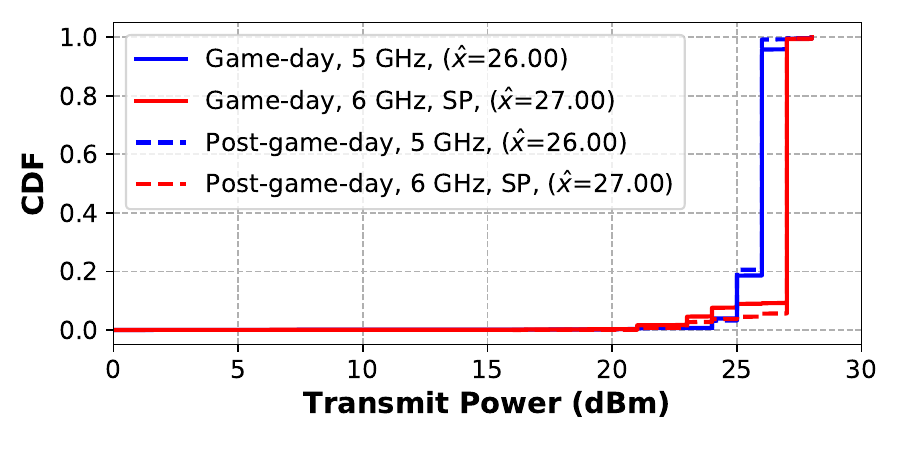}
    \caption{}
    \label{fig:outdoor_txp_5v6}
    \end{subfigure}
    \begin{subfigure}{\linewidth}
\centering\includegraphics[width=\linewidth]{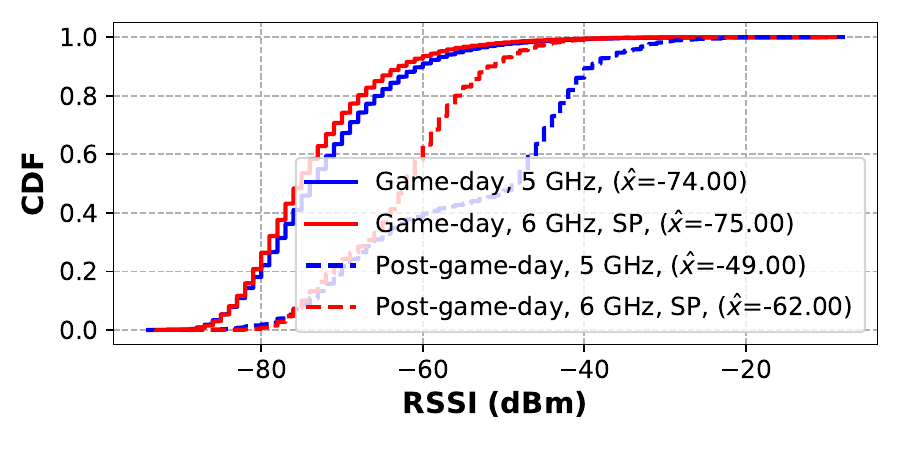}
    \caption{}
    \label{fig:outdoor_rssi_5v6}
    \end{subfigure}
    % \vspace{-.5em}
    \caption{Tx power and RSSI levels for 5 GHz and 6 GHz SP outdoors on game-day.}
    \label{fig:outdoor_5v6}
    \vspace{-1em}
 \end{figure}
 
%%%%%%%%%%%%%%%%%%%%%%%%%%%%%%%%%%%%%%%%%% 
\section{Results \& Discussions}
The vast amount of data collected on game-day and post-game-day was carefully analyzed to derive insights into coexistence in the 6 GHz band. In this section, we present our analyses of (i) Wi-Fi usage within the stadium, (ii) coexistence of outdoor SP and indoor LPI deployments, and (iii) analysis of spectrum analyzer measurements. It should be noted that all SigCap measurements of Wi-Fi parameters were collected with only the six phones listed in Table~\ref{tab:FeatureDevices}.

\subsection{ Wi-Fi Usage in The Stadium: 5 GHz vs. 6 GHz}\label{sec:usage}

Each AP deployed in the stadium broadcasts ND-guest on game-day and post-game-day, on both 5 GHz and 6 GHz, allowing easy comparison of the usage of the two bands on this SSID. Fig. \ref{fig:outdoor_5v6} shows \chk{empirical CDFs, comparing the distribution of} Tx power and measured RSSI levels for 5 GHz and 6 GHz SP APs outdoors, on game-day and post-game-day, \chk{as well as the median of data ($\hat{x}$). In particular, the Tx power parameter is extracted from Wi-Fi beacons, and the 802.11 specification defines this as the transmission power of the AP with $\pm$5 dB tolerance to the real EIRP. In the SP Wi-Fi 6E APs, we observe a median Tx power of 27 dBm over 80 MHz, while the median Tx power for 5 GHz is 26 dBm over 20 MHz, as seen in Fig. \ref{fig:outdoor_txp_5v6}. We observe this on both game-day and post-game-day. This is lower than the maximum 36 dBm EIRP permitted, both by the AFC in 6 GHz and by FCC regulations in 5 GHz.}
Since the deployment is extremely dense, lower Tx power allows better frequency reuse. Fig. \ref{fig:outdoor_rssi_5v6} shows the measured RSSI over the 20 MHz beacons: we notice a significant increase in measured RSSI on both bands on post-game-day, most likely due to the stadium being empty leading to reduced body-loss.

% As described in Section \ref{sec:deployments}, both 5 and 6 GHz transmissions are generated from the same unit. SP operation in the 6 GHz band is comparable to the existing FCC rules for the 5 GHz spectrum: the key difference is the introduction of a maximum Power Spectral Density (PSD) limit of 23 dBm/MHz in the 6 GHz band. In Fig. \ref{fig:outdoor_txp_5v6}, we observe similar median transmit (Tx) power levels over different bandwidth, with 26 dBm over 40 MHz for 5 GHz  nd 27 dBm over 80 MHz for 6 GHz. Fig. \ref{fig:outdoor_rssi_5v6} also shows similar outdoor RSSI levels for 5 GHz and 6 GHz SP APs, indicating they can provide comparable link performance for STAs. 

\begin{figure}
 \centering
    \begin{subfigure}{\linewidth}
    \centering
    \includegraphics[width=\linewidth]{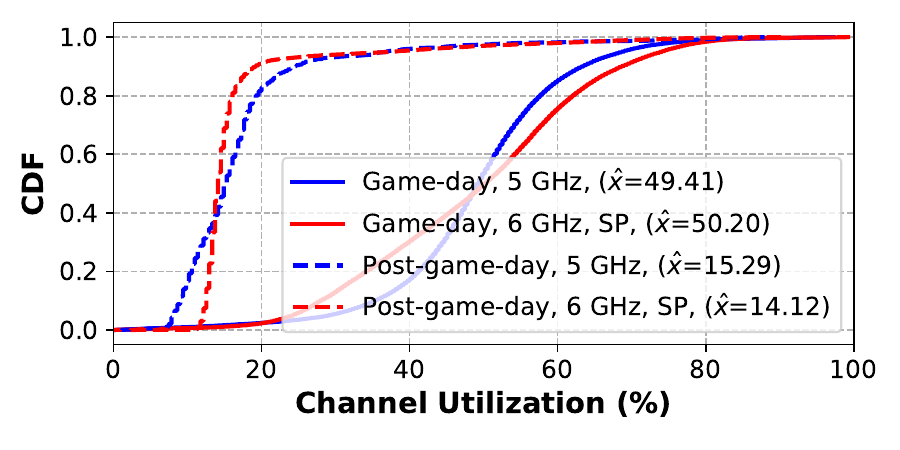}
    \caption{}
    \label{fig:cutil5v6}
    \end{subfigure}
    \begin{subfigure}{\linewidth}
    \centering
    \includegraphics[width=\linewidth]{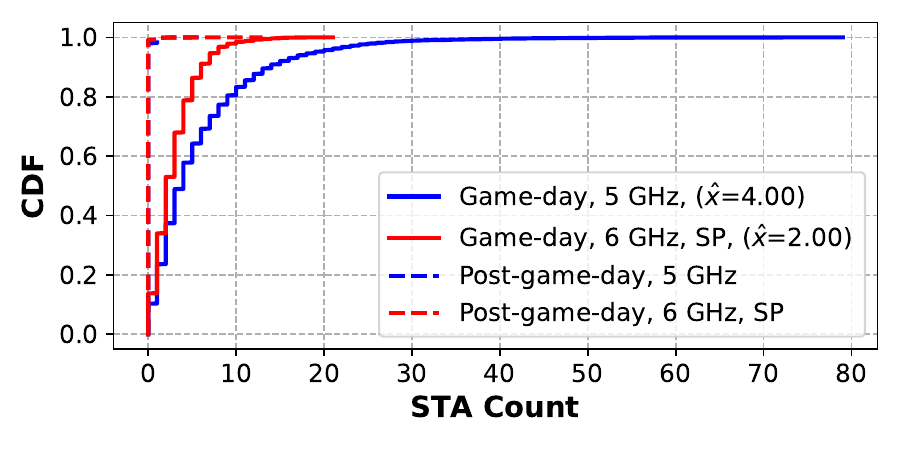}
    \caption{}
    \label{fig:sta5v6}
    \end{subfigure}
%      \begin{subfigure}{\linewidth}
%     \centering
% \includegraphics[width=\linewidth]{figures/cdf-stacount-5-6-v2.pdf}
%     \caption{}
%     \label{fig:sta5v6}
%     \end{subfigure}
    % \vspace{-.5em}
    \caption{ Channel utilization and STA count for 5 GHz vs. 6 GHz SP outdoors on game-day and post-game-day.}
    \label{fig:outdoor_5v6_2}
    \vspace{-1em}
 \end{figure}

Fig. \ref{fig:outdoor_5v6_2} shows the CDFs of primary channel utilization and station (STA) count at each timestamp. In Fig. \ref{fig:cutil5v6}, we see similar channel utilization for 5 GHz and 6 GHz SP APs, on both game-day and post-game-day, since primary channel utilization is mainly due to the transmission of beacons, which is similar in both bands. Although the STA count for 6 GHz in Fig. \ref{fig:sta5v6} remains lower than that of 5 GHz,  we observe a significant increase compared to the results presented in \cite{dogan2023evaluating} from measurements in 2023. The lower STA count for 6 GHz is partly due to the fact that not all phones in the stadium were 6 GHz-capable, limiting the number of devices connecting to the 6 GHz band. We also see that the STA count on the post-game day is substantially lower than on game-day, as expected.

\begin{figure}
    \includegraphics[width=\linewidth]{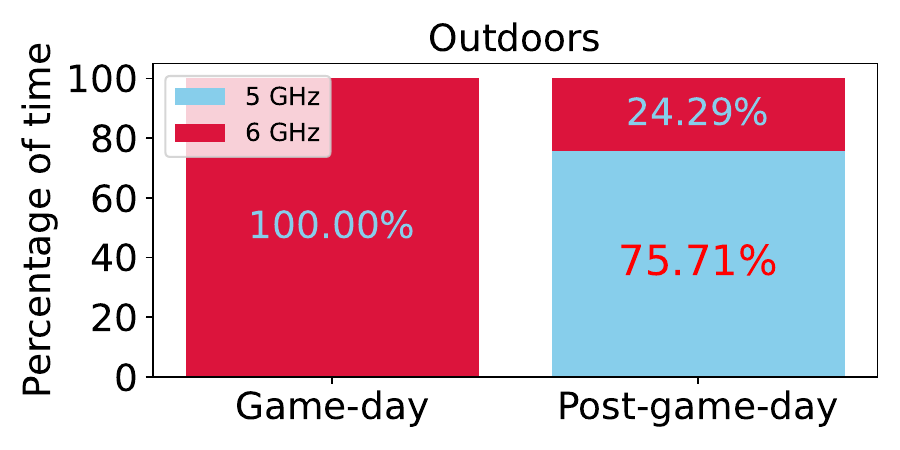}
    % \vspace{-1.5em}
    \caption{Wi-Fi band preference of the connected measurement phones on game-day, and post-game-day.}
    \label{fig:bar_connected}
    \vspace{-1em}
\end{figure}

\begin{figure}
    \includegraphics[width=\linewidth]{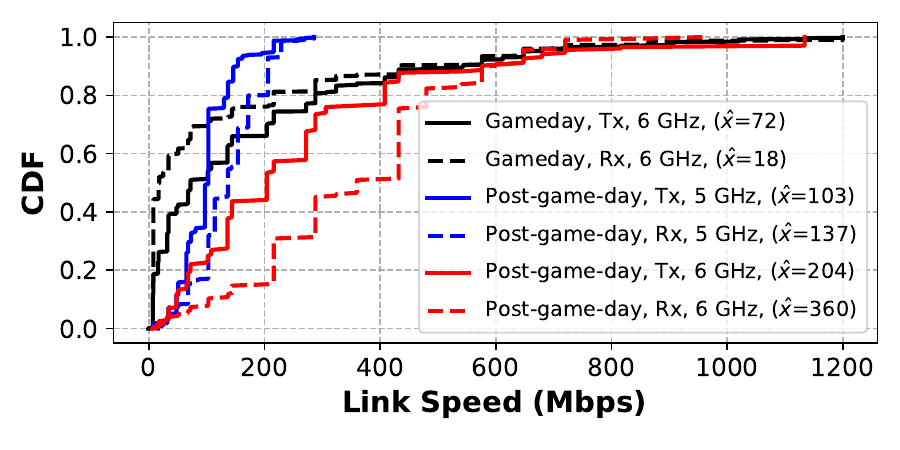}
    \caption{Data rates for the measurements in Fig. \ref{fig:bar_connected}.}
    \label{fig:cdf_tx_rx_link_speed}
    \vspace{-1em}
\end{figure}

\begin{figure}
    \includegraphics[width=\linewidth]{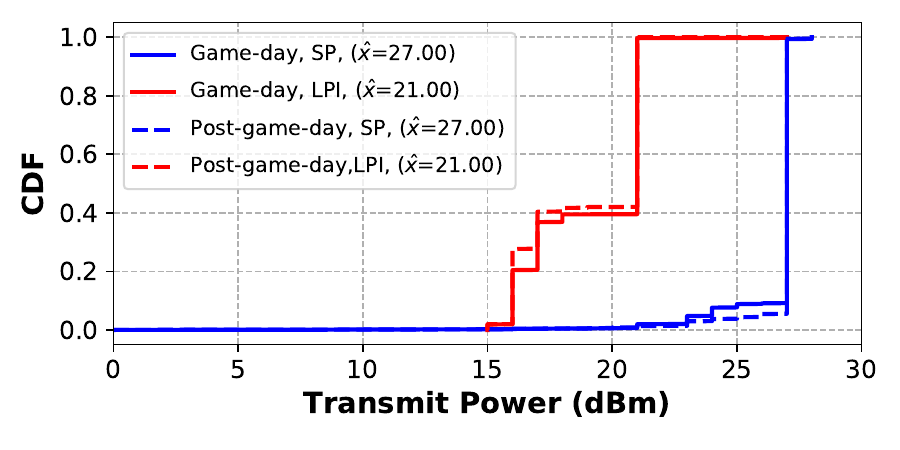}
    \caption{Transmit power for SP and LPI APs.}
    \label{fig:cdf_dp_lp_tx}
\end{figure}

Fig. \ref{fig:bar_connected} illustrates the connection ratios of the Wi-Fi bands on game-day and post-game-day. It is important to note that this analysis is based on the connection status of only the measurement phones used during the measurements, not all phones present in the stadium during game-day (80k).
On game-day, the measurement phones were always connected to Wi-Fi 6E SP APs, with no connections to 5 GHz APs. This is likely because the 5 GHz band was fully utilized, as also indicated by the STA count in Fig. \ref{fig:sta5v6} . However, during the post-game-day measurement, the phones were able to freely choose between the 5 GHz and 6 GHz bands, as the stadium was empty and 75\% of the connections were to 5 GHz. This result indicates that the 6 GHz band is crucial for improving connectivity in dense deployments: the 5 GHz band is almost completely saturated on game-day. In addition to these measurements, according to data provided to us from the back-end of the deployment by the Notre Dame Office of Information Technologies, 14\% of client connections in the stadium on game-day were over the 6 GHz band: this rate of adoption is fairly high since only the newer phone models incorporate Wi-Fi 6E.  

Since our research focus for this paper is on evaluating spectrum coexistence behavior, throughput tests were not conducted. However, we analyzed the Tx and Rx link-speed values broadcast by each beacon, which represent uplink and downlink connected link-speeds respectively.  Fig. \ref{fig:cdf_tx_rx_link_speed} shows these link-speeds for the same data-set analyzed in Fig. \ref{fig:bar_connected}.  Since the measurement phones were only connected to 6 GHz on game-day, there are no results for 5 GHz, but comparing game-day and post-game-day measurements for 6 GHz, we see that game-day link speeds are substantially lower, particularly on downlink. For the post-game-day comparison between 5 GHz and 6 GHz, there is a substantial difference between speeds over 20 MHz channels in 5 GHz compared to 80 MHz channels over 6 GHz, with the maximum link speed reaching up to 1200 Mbps for 6 GHz, while it is only around 300 Mbps for 5 GHz.
% Transmission over 80 MHz at 6 GHz provides higher Tx/Rx data rates compared to 5 GHz. We observe a significant difference in the maximum achievable data rates, with the maximum link speed reaching up to 1200 Mbps for 6 GHz, while it is around 300 Mbps for 5 GHz. ... 

% Six phones given in Table \ref{tab:FeatureDevices} were used to collect measurements in different regions of the stadium. While these measurements provide valuable insights into the overall situation, they do not offer a complete view of client connections across the entire stadium. According to data from the Office of Information Technologies, 16\% of client connections in the stadium is over 6 GHz band, just two months after the outdoor deployment began.

% On gameday, only a small fraction of the data samples collected on the six phones were connected to any Wi-Fi, accounting for approximately 1.2\% (1,866 out of 158,966 samples), and all of these connections were to Wi-Fi 6E SP APs, with no connections to 5 GHz APs. This is likely because the 5 GHz band was fully utilized. However, during the post-game-day measurement, the six phones were able to freely choose between the 5 GHz and 6 GHz bands, as the stadium was empty. %The high connection ratios to 6 GHz in INW (67.5 \%) and II (74.73\%) environments on postgameday indicate that 6 GHz-capable phones prefer to connect to 6 GHz rather than 5 GHz.

% GD connections 
% for outdoors: 1866 all SPs

\color{black}

%  \begin{figure}
% \begin{subfigure}{0.49\linewidth}
%     \includegraphics[width=\linewidth]{figures/bar-connected-5-6GHz-all.pdf}
%     \caption{}
%     \label{fig:bar_connected_game}
%     \end{subfigure}
%      \hfill
%     \begin{subfigure}{0.49\linewidth}
%     \includegraphics[width=\linewidth]{figures/bar-connected-5-6GHz-all_postgame.pdf}
%     \caption{}
%     \label{fig:bar_connected_postgame}
%     \end{subfigure}
%     % \vspace{-1.5em}
%     \caption{Wi-Fi band preference of the measurement phones on gameday, and post-gameday.}
%     \label{fig:bar_connected}
% \end{figure}

\subsection{Coexistence of outdoor SP and indoor LPI deployments} \label{sec:impact_:sp_lpi}

It is important to note that while SP APs can also be deployed indoors under AFC control, in the deployment we studied, SP APs are only deployed outdoors in the bowl. However signals from outdoor SP APs propagate indoors and those from indoor LPI APs propagate outdoors. In this section we carefully analyze these interactions.

Fig. \ref{fig:cdf_dp_lp_tx} shows transmit power levels for SP and LPI APs, \chk{as reported in the beacons}. We see that both levels are lower than the maximum permitted for 80 MHz channels, which are 36 dBm and 24 dBm respectively. This is not uncommon in dense enterprise deployments since frequency reuse is enabled by lower transmitted power. Lower transmit power is also beneficial for coexistence with incumbents and between outdoor and indoor deployments.

\begin{figure}
    \begin{subfigure}{0.49\linewidth}
    \includegraphics[width=\linewidth]{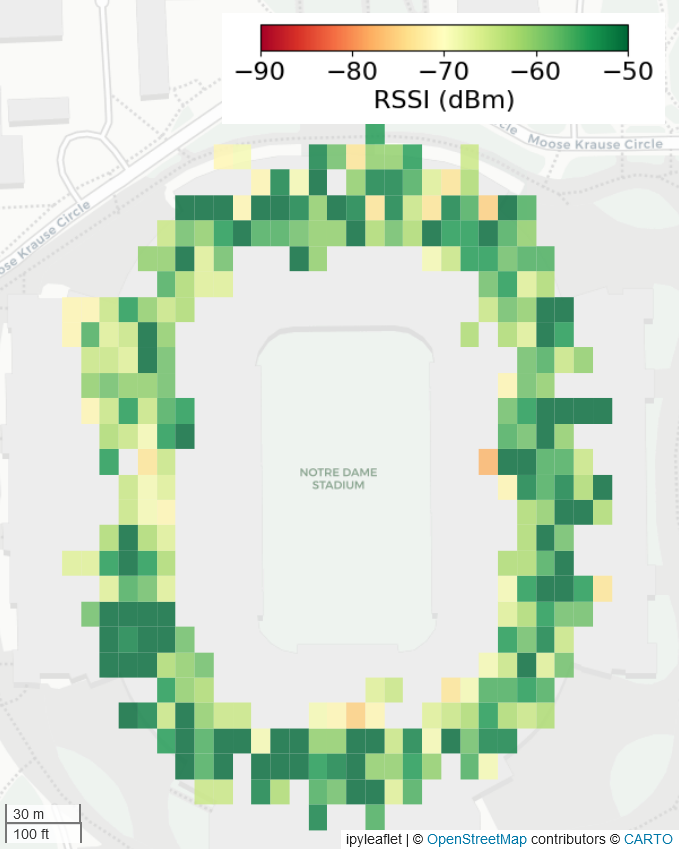}
    \caption{SP APs}
    \label{fig:heatmap_sp_bowl}
    \end{subfigure}
    \begin{subfigure}{0.49\linewidth}
    \includegraphics[width=\linewidth]{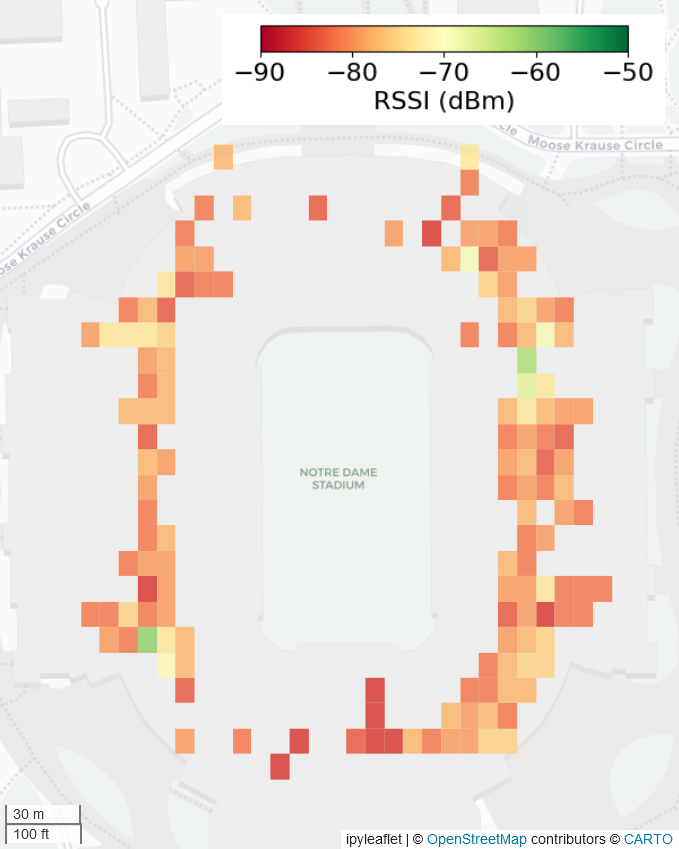}
    \caption{LPI APs}
    \label{fig:heatmap_lpi_bowl}
    \end{subfigure}
    % \vspace{-1.5em}
    \caption{Outdoor RSSI heatmap of SP APs and LPI APs on game-day.}
    \label{fig:heatmap_outdoors}
\end{figure}

Fig. \ref{fig:heatmap_outdoors} shows the outdoor RSSI heatmap of SP APs and LPI APs on game-day in the bowl. 
% Walking measurements during the game were restricted due to the stadium being at full capacity and hence measurements were conducted walking up and down the stairs and walkways between tiers. 
As seen in Fig. \ref{fig:heatmap_sp_bowl}, the SP APs provide consistent and strong coverage across the stadium, with RSSI levels ranging from -70 dBm to -50 dBm. In Fig. \ref{fig:heatmap_lpi_bowl}, we see that LPI APs (deployed indoors) are also measurable outdoors but at significantly lower RSSI compared to the SP APs: this is due to the lower density of LPI APs, lower Tx power (Fig. \ref{fig:cdf_dp_lp_tx}) and many of the LPI APs being in indoor interior environments with greater isolation from the outdoors.

% and provide localized coverage with lower RSSI values on the left and right sides of the stadium, where the measurement buildings with LPI APs have large windows facing the stadium as shown in Fig. \ref{fig:NDSmeasurment}.

\begin{figure}
    \begin{subfigure}{\linewidth}
    \includegraphics[width=\linewidth]{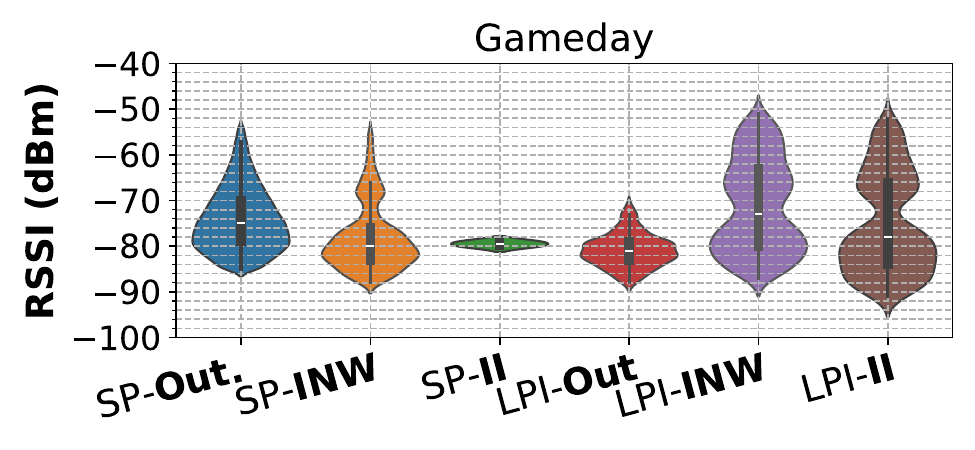}
    \caption{}
    \label{fig:rssi_gameday}
    \end{subfigure}
    \hfill
    \begin{subfigure}{\linewidth}
    \includegraphics[width=\linewidth]{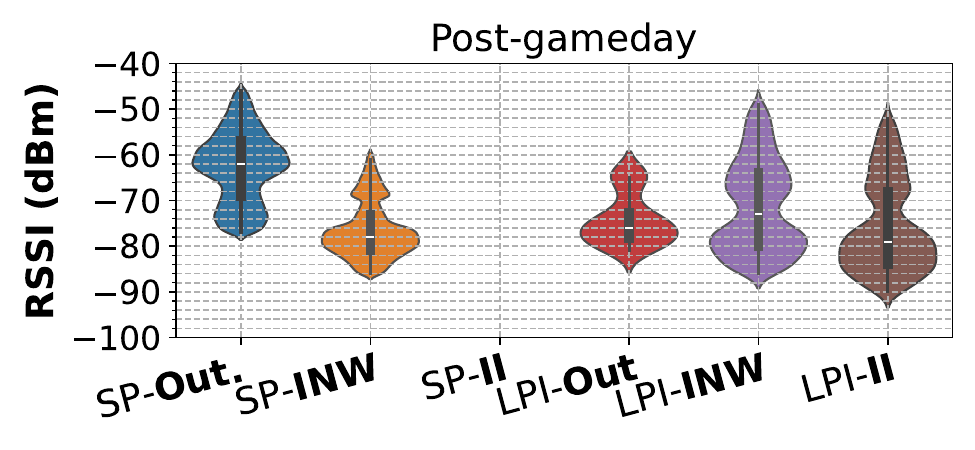}
    \caption{}
    \label{fig:rssi_postgame}
    \end{subfigure}
    % \vspace{-1.5em}
    \caption{Measured RSSI levels on game-day and post-game-day. SP/LPI-Out: measured outdoors from SP/LPI APs, SP/LPI-INW: measured indoors near windows from SP/LPI APs, SP/LPI-II: measured indoors interior from SP/LPI APs.} 
    \label{fig:RSSI}
\end{figure}

\begin{figure}
    \begin{subfigure}{0.95\linewidth}
    \includegraphics[width=\linewidth]{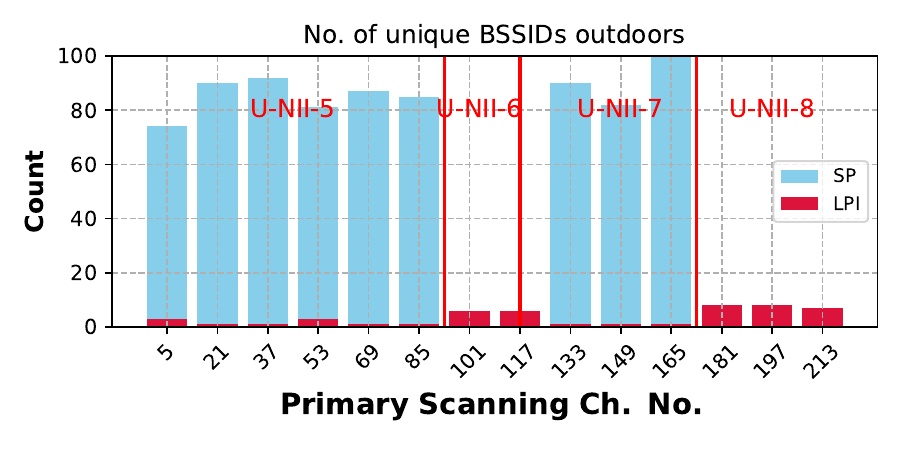}
    \caption{}
    \label{fig:bar_bssid_bowl}
    \end{subfigure}
    %\hfill
    \begin{subfigure}{0.95\linewidth}
    \includegraphics[width=\linewidth]{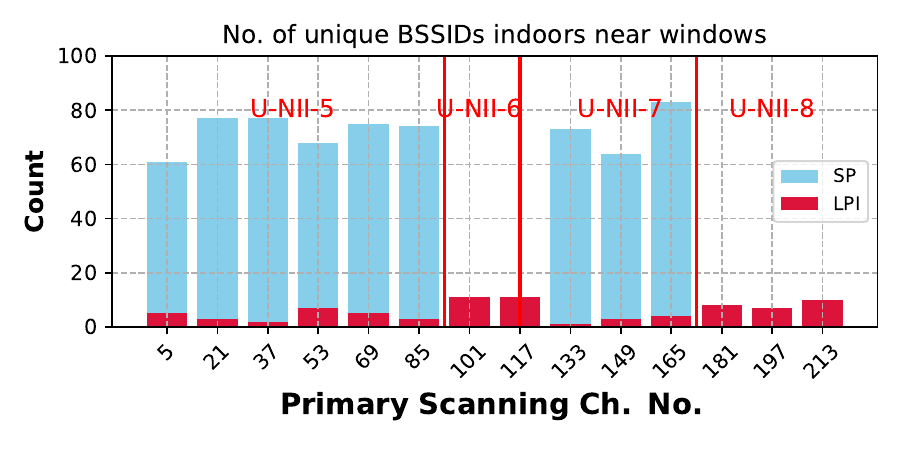}
    \caption{}
    \label{fig:bar_bssid_corbett789}
    \end{subfigure}
    \begin{subfigure}{\linewidth}
    \includegraphics[width=0.95\linewidth]{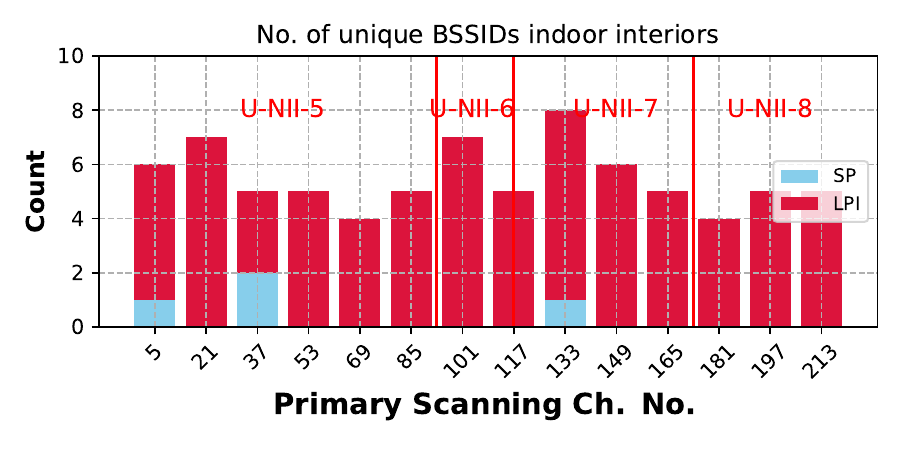}
    \caption{}
    \label{fig:bar_bssid_duncan12}
    \end{subfigure}
    % \vspace{-1.5em}
    \caption{No. of unique BSSIDs on game-day for outdoors (Out), indoors near windows (INW) and indoor interior (II).}
    \label{fig:bar_bssid_gameday}
\end{figure}

\begin{figure}
    \begin{subfigure}{0.95\linewidth}
    \includegraphics[width=\linewidth]{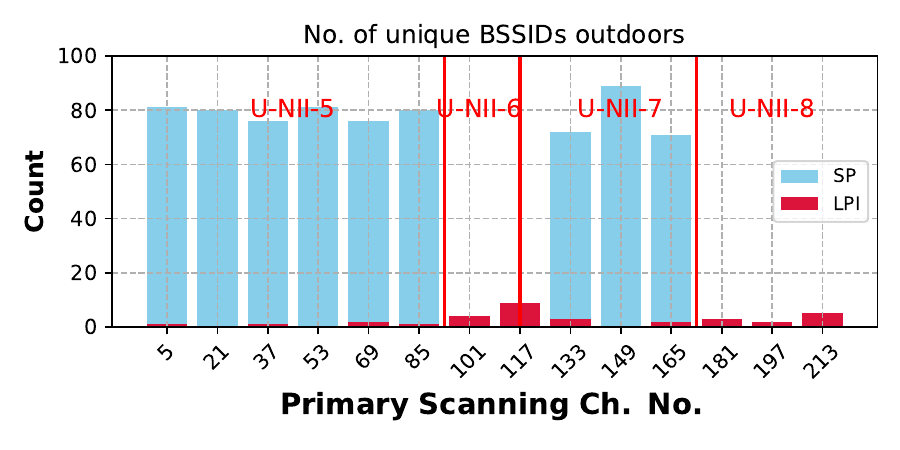}
    \caption{}
    \label{fig:bar_bssid_bowl_postgame}
    \end{subfigure}
    \begin{subfigure}{0.95\linewidth}
    \includegraphics[width=\linewidth]{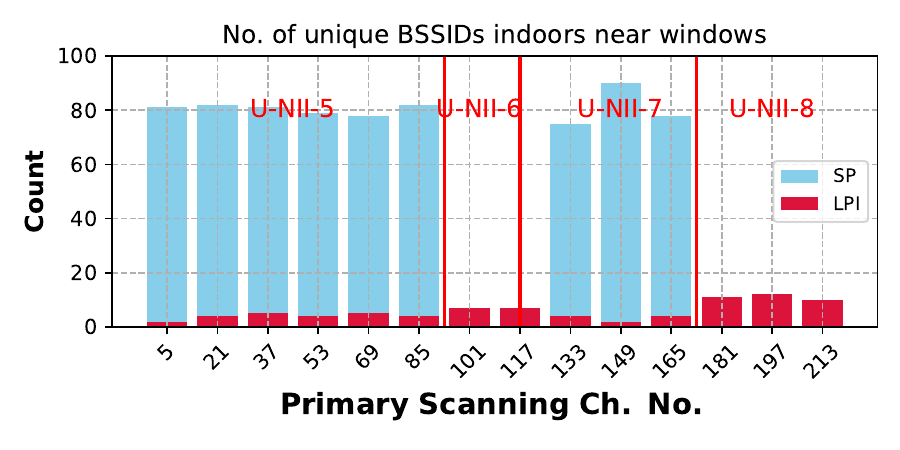}
    \caption{}
    \label{fig:bar_bssid_corbett789_postgame}
    \end{subfigure}
     \begin{subfigure}{0.95\linewidth}
    \includegraphics[width=\linewidth]{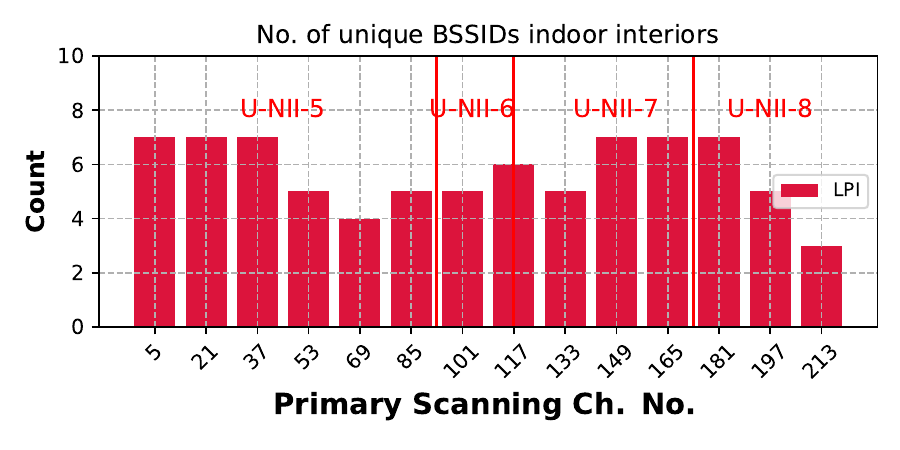}
    \caption{}
    \label{fig:bar_bssid_duncan12_postgame}
    \end{subfigure}
    % \vspace{-1.5em}
    \caption{No of unique BSSIDs on post-game-day for outdoors (Out), indoors near windows (INW) and indoor interior (II).}
    \label{fig:bar_bssid_postgame}
\end{figure}

\begin{figure} 
    \begin{subfigure}{0.85\linewidth}
    \centering
    \includegraphics[width=\linewidth]{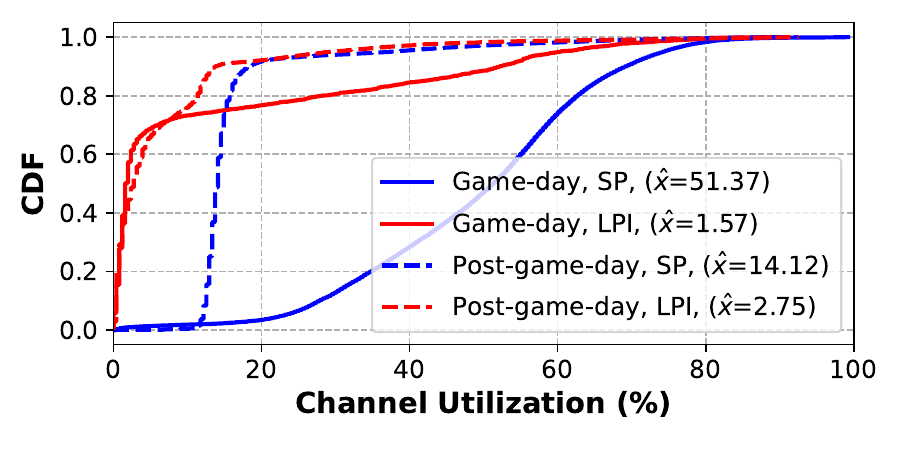} 
    \caption{}
    \label{fig:ch_util_sp_game}
    \end{subfigure}
    %\hfill
    \begin{subfigure}{0.85\linewidth}
    \centering
   \includegraphics[width=\linewidth]{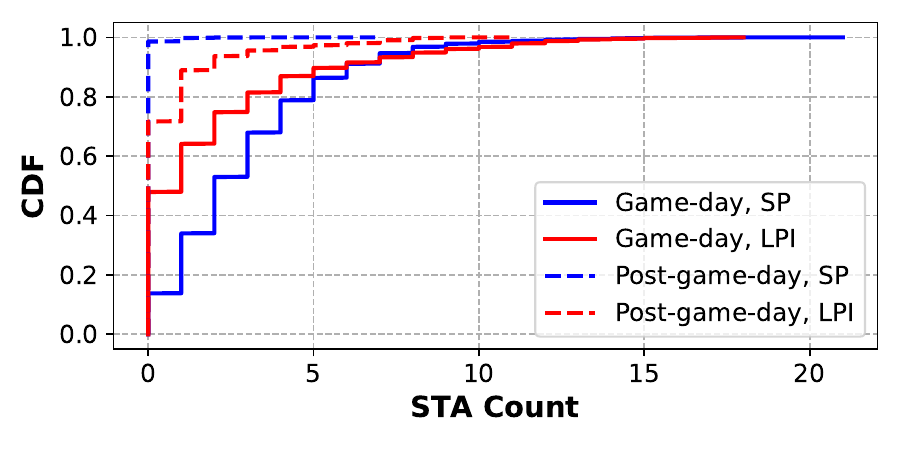} 
    \caption{}
    \label{fig:stacount_sp_game}
    \end{subfigure}
    % \vspace{-1.5em}
    \caption{Channel utilization and STA count for SP and LPI on game-day and post-game-day.}
    \label{fig:ch_util_sta_count_game}
\end{figure}

% \begin{figure}
%     \begin{subfigure}{0.85\linewidth}
%     \centering
%      \includegraphics[width=\linewidth]{figures/cdf-chutil-sp-postgameday.pdf}  
%     \caption{}
%     \label{fig:ch_util_sp_postgame}
%     \end{subfigure}
%     % \hfill
%     \begin{subfigure}{0.85\linewidth}
%     \centering
%     \includegraphics[width=\linewidth]{figures/cdf-stacount-postgameday.pdf} 
%     \caption{}
%     \label{fig:stacount_sp_game}
%     \end{subfigure}
%     % \vspace{-1.5em}
%     \caption{Channel utilization and STA count for SP outdoors and LPI indoors on gameday}
%     \label{fig:ch_util_sta_count_postgame}
% \end{figure}

Fig. \ref{fig:RSSI} shows beacon RSSI levels over the 20 MHz primary scanning channels measured during walking measurements on game-day and post-game-day across the three environments.
\chk{In particular, Fig. \ref{fig:rssi_gameday} exhibit comparable RSSI levels during game-day in SP APs measured indoors near windows (SP-INW) and LPI APs measured indoors near windows (LPI-INW), with a median difference of approximately 7 dB. This observation is consistent on both game-day and post-game-day as shown in Fig. \ref{fig:rssi_postgame}, with a lower median difference of 5 dB. This suggests that both SP and LPI APs share similar characteristics in the INW environment, \ie SP operations outdoors and LPI operations indoors will affect each other, further highlighting the importance of careful channel assignments in the INW environment.}
% Fig. \ref{fig:rssi_gameday} exhibit similar RSSI levels during dame-day in SP APs measured outdoors (SP-Out) and SP APs measured indoors near windows (SP-INW). Additionally, SP-INW RSSI levels are comparable to those of LPI measured indoors near windows (LPI-INW), indicating that both SP and LPI share similar characteristics in the INW environment. These results show that SP operations outdoors and LPI operations indoors will affect each other, highlighting the importance of careful channel assignments in the INW environment.
On the other hand, we observe low RSSI levels for SP APs measured in indoor interior environments (SP-II). The measurements indicate that indoor interior environments, with their physical separation, are well insulated from outdoor SP operations. This isolation enables both outdoor and indoor interior deployments to operate simultaneously without negatively impacting each other's performance.
When comparing Figs. \ref{fig:rssi_gameday} and \ref{fig:rssi_postgame}, we observe \chk{that the median RSSI from SP APs measured outdoors (SP-Out) is 13 dB higher on post-game-day, than on game-day: possibly due to increased signal attenuation from body-loss when the stadium is at full capacity.}
% a 10 dB increase in the median RSSI level measured outdoors during the post-game-day measurements: this could be due to increased signal attenuation due to body-loss when the stadium is at full capacity.

In Fig.~\ref{fig:bar_bssid_gameday} and Fig.~\ref{fig:bar_bssid_postgame} we show the number of unique BSSIDs across the U-NII bands measured on game-day and post-game-day, respectively, in the three environments. A total of 785 unique BSSIDs were observed for SP APs in the bowl across nine 80 MHz channels, with the majority showing a similar number of unique BSSIDs, as shown in  Fig. \ref{fig:bar_bssid_bowl}.

Furthermore, comparing Fig.~\ref{fig:bar_bssid_bowl} and Fig.~\ref{fig:bar_bssid_corbett789} we see that the indoors near windows environment is very similar to outdoors, whereas we see from Fig.~\ref{fig:bar_bssid_duncan12} that the indoor interior environment is quite isolated from the outdoors since almost no beacons from SP APs are received there. A similar pattern is observed in the post-game-day measurements shown in Fig.~\ref{fig:bar_bssid_postgame}. Hence, SP APs will affect LPI APs unless they are isolated from each other by significant building loss.

\color{black}

Fig. \ref{fig:ch_util_sta_count_game} shows the CDF plots of channel utilization and STA count on game-day and post-game-day. Clearly, the SP APs deployed outdoors in the bowl are used more intensely during the game-day (STA count on post-game-day is very low) whereas the LPI AP usage on the post-game-day decreases only slightly: since these are deployed indoors and post-game-day was a working day, it is likely that there were multiple clients connected to them indoors. 
% The STA count and channel utilization metrics are broadcast by the AP in the beacon and remains consistent whether the beacon is received indoors or outdoors. n ofFig. \ref{fig:stacount_sp_game} shows STA count ...

Given the observation of similar RSSI levels for SP and LPI measured INW in Fig. \ref{fig:RSSI}, Table \ref{tab:unii_band_ratios} presents a detailed analysis of 6 GHz connections in this environment. On game-day, 29.24\% of 6 GHz connections were made to a SP AP, while 70.76\% were to the LPI AP. In contrast, on post-game-day, the percentage of connections to outdoor SP APs increases to 69.1\% in INW environments, since the RSSI is comparable to LPI APs installed indoors but the stadium is empty and the SP APs are not being utilized by outdoor clients. Moreover we also see that on game-day there were more connections to U-NII-6 and U-NII-8 (32.5\%) compared to post-game-day (27.7\%) indicating that dense outdoor usage on the SP bands (U-NII-5 and U-NII-7) will affect usage of indoor LPI APs since indoor clients will preferentially connect to the LPI-only bands in U-NII-6 and U-NII-8 when outdoor SP usage in U-NII-5 and U-NII-7 increases. Our conclusion from this analysis is that (i) similar RSSI levels from SP APs measured in both INW and outdoor environments enable indoor phones to connect to the outdoor SP APs, and (ii) the degree of usage of outdoor networks will impact connections within INW environments. Since both deployments use the same Wi-Fi coexistence protocol, Carrier Sense Multiple Access with Collision Avoidance (CSMA/CA), this adaptation occurs seamlessly without the need for additional signaling, but this may not be true for dissimilar systems.

% Moreover, we see that U-NII-5 usage by LPI APs is the lowest (4.4\%) because it is mostly used by SP APs outdoors, leading to high channel utilization as expained in Fig. \ref{fig:ch_util_sp_game}. U-NII-7 is the most commonly used, with a connection ratio of 63\%. This can be explained via the lowest channel utilization observed indoors. Also, channel 117 in U-NII-7 is avoided by outdoors SP APs because it overlaps with U-NII-6

%\color{red}
%During the gameday measurements, phones were connected to SP APs 29.24\% of the time, while this increased to 69.1\% for post-gameday measurements. The higher post-gameday ratio is due to the stronger signal received from SP APs, which led phones to connect to them more often.

%Looking at the gameday results, the connected LPI data shows that LPI APs were predominantly connected to U-NII-7. Based on the channel utilization plots in the paper, we observe that LPI APs experience higher channel utilization on U-NII-5 due to SP operations outdoors. 

%As a result, LPI APs tend to shift to other U-NII bands. U-NII-6 and U-NII-8 were used 32.5\% of the time during gameday, which decreases to 27.7\% in post-gameday measurements.

\begin{table}
\centering
\renewcommand{\arraystretch}{1.2}
\caption{Connection ratios for SP and LPI for indoors near windows (INW) environment on game-day and post-game-day.}
\begin{tabular}{|c||c|c|}
\hline
\multirow{2}{*}{\textbf{Meas}} & \multicolumn{2}{c|}{\textbf{ INW 6 GHz Connections}} \\ \cline{2-3}
 & {\textbf{to SP AP}} & \textbf{to LPI AP} \\ \hline  \hline
\multirow{4}{*}{\textbf{Game-day}} 
 &  & 70.76\% \\ 
 & \multicolumn{1}{c|}{29.24\%} 
 & \multicolumn{1}{c|}{\textbf{UNII-Band Ratios:}} \\ 
 & & U-NII-5 and U-NII-7: 67.5\% \\ 
 & & U-NII-6 and U-NII-8: 32.5\% \\  \hline
\multirow{4}{*}{\textbf{Post-game-day}} 
 & & 30.9\% \\
 & \multicolumn{1}{c|}{69.1\% } 
 & \multicolumn{1}{c|}{\textbf{UNII-Band Ratios:}} \\ 
 &&  U-NII-5 and U-NII-7: 72.3\% \\ 
 &  & U-NII-6 and U-NII-8: 27.7\% \\ 
 \hline
\end{tabular}

\label{tab:unii_band_ratios}
\end{table}

\subsection{Aggregate interference in 6 GHz}\label{sec:indoor_outdoor_rsrp}
%\subsection{Aggregate interference \& fixed links}\label{sec:indoor_outdoor_rsrp}
In normal Wi-Fi deployments, aggregate interference is usually not a concern since APs in close proximity on the same channel will defer to each other and not transmit simultaneously. However, in an extremely dense deployment, such as the ND stadium with approximately 100 BSSIDs broadcasting on each channel, aggregate interference needs to be studied. However, unlike the CBRS SAS, the 6 GHz AFC calculations do not consider aggregate interference when allocating power, even in dense deployments. 

\begin{figure}
    \begin{subfigure}{\linewidth}
    \includegraphics[width=\linewidth]{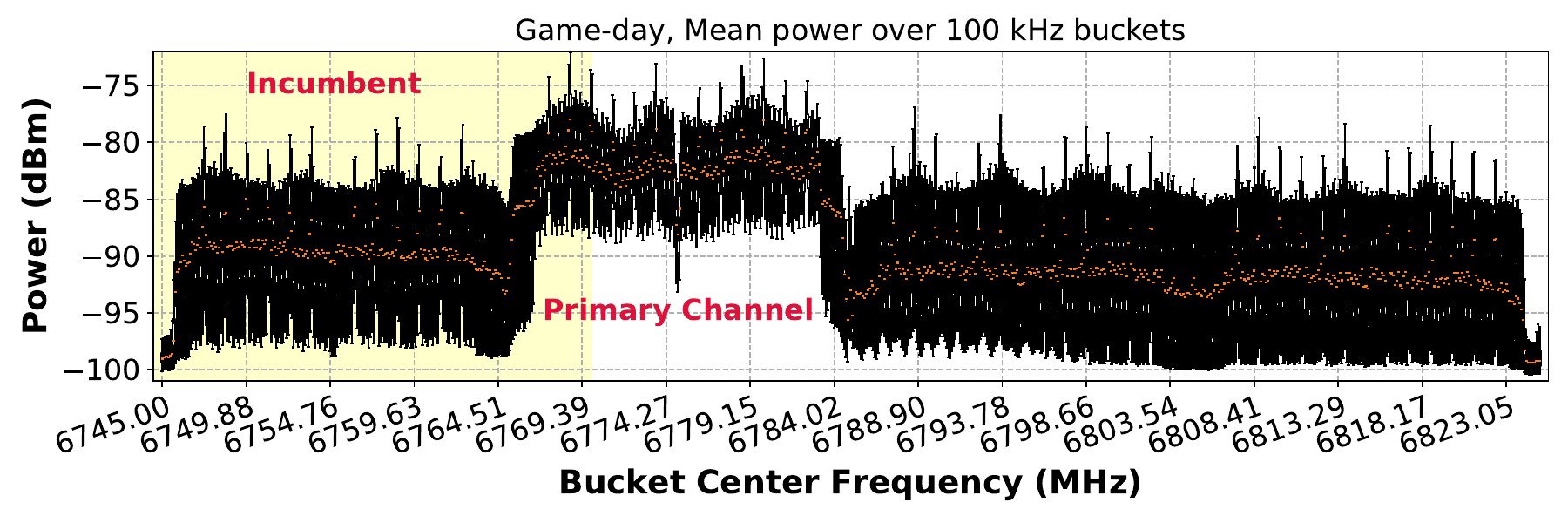}
    \caption{}
    \label{fig:spectrum_gameday_2}
    \end{subfigure}
    \hfill
    \begin{subfigure}{\linewidth}
    \includegraphics[width=\linewidth]{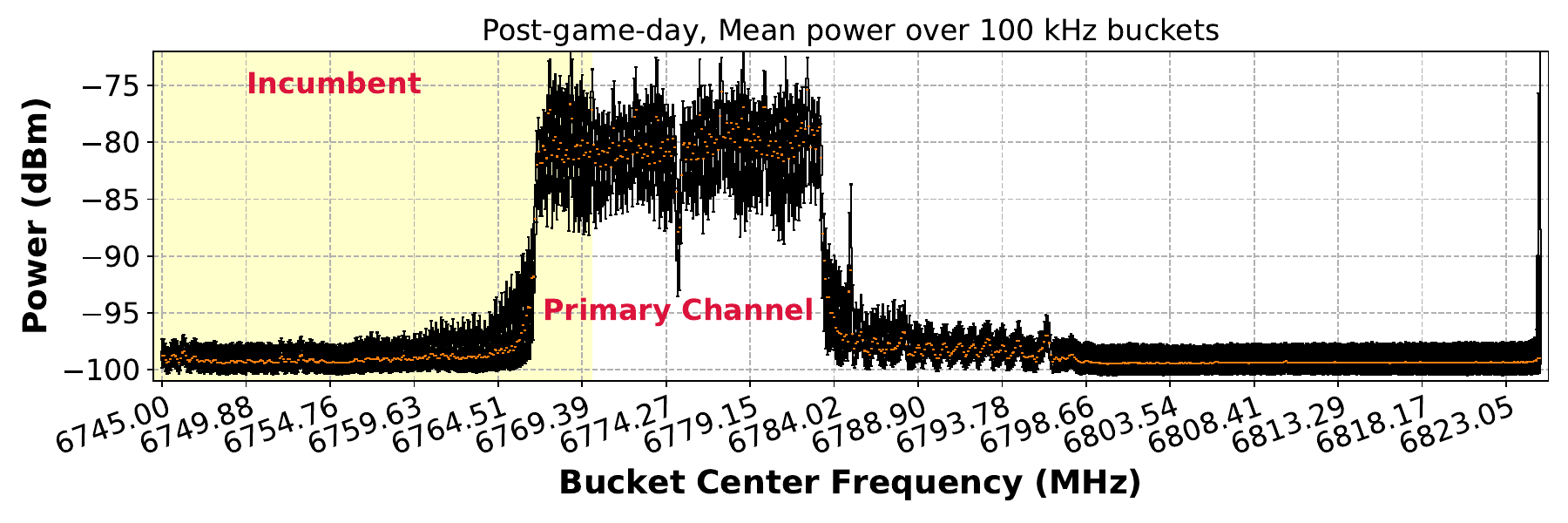}
    \caption{}
    \label{fig:spectrum_postgame_2}
    \end{subfigure}
    % \vspace{-1.5em}
    \caption{Spectrum analyzer captures on game-day and post-game-day on Wi-Fi Channel 167 (6745 - 6825 MHz), with incumbent on 6740 - 6770 MHz. } 
    \label{fig:spectrum_gd_pgd}
\end{figure}

Spectrum analyzer measurements were taken over all fourteen 80 MHz channels on game-day and post-game-day, about 2000 captures for each channel spread over a few hours. These exhibit similar characteristics and hence we present results for a particular channel that has a coexisting incumbent on a portion of the channel.
Fig.~\ref{fig:spectrum_gd_pgd} shows the individual captures as well as the mean power on Channel 167 (6745 - 6825 MHz). The 30 MHz incumbent with center frequency 6755 MHz is also shown in yellow in the figure, overlapping part of the primary channel and part of the data channel. The 20 MHz primary channel, Channel 165 (6765 - 6785 MHz) has an elevated power level due to beacon transmissions: this does not differ between game-day and post-game-day measurements, which is to be expected since beacon traffic is independent of number of users.

\chk{AFC calculates the allowed EIRP for an SP AP by considering the AP's location, nearby incumbent locations, and the incumbents' operating frequencies. From those information, it calculates ratio of interference to noise power (I/N ratio) caused by the AP at the incumbents for all Wi-Fi channels in 6 GHz. The AFC limits an AP's EIRP such that it does not cause a harmful interference, which is defined as I/N of -6 dB. However, AFC systems as mandated by the FCC do not require aggregate interference taken into consideration~\cite{FCC2}.
Our observation shows a significantly higher aggregate interference measured on game-day: 10 dB higher in mean power and 20 dB higher in peak power compared to post-game-day. This elevated interference levels on game-day, which are caused by sustained traffic between APs and users in a dense scenario, may cause interference to a nearby incumbent.}
% However, there is a significant difference in aggregate interference outside the primary channel bandwidth, between game-day and post-game-day, when the APs are actively connected to users. We see an increase of about 10 dB in mean power and about 20 dB in peak power: these levels may cause interference to a nearby incumbent since \chk{AFC calculates allowed power by assuming only one AP is actively transmitting.}
In this particular location, the stadium was not in the main beam of the incumbent receiver and hence it is unlikely that interference occurred, but there may be other, similarly dense outdoor deployments that are in the path of receive beams of incumbent 6 GHz links.

%%%%%%%%%%%%%%%%%%%%%%%%%%%%%%%%%%%%%%%%%%%%%%%%%%%%%

%\subsection{Comparison of Indoor-to-Outdoor and Outdoor-to-Indoor Losses}\label{sec:ito_oti}

% \begin{figure}
%     \centering
%     \includegraphics[width=\linewidth]{figures/cdf_throughput_combined_inw_out_ordered.png}
%     \caption{CDF of throughput by U-NII band for INW and OUT connections.}
%     \label{fig:unii_connection}
% \end{figure}

% The throughput is obtained based on wifi-connected-tx-link-speed column in general CSV files.

%%%%%%%%%%%%%%%%%%%%%%%%%%%%%%%%%%%%%%%%%% 
% \vspace{-1em}
\section{Conclusions \& future research}
The first-of-its-kind, comprehensive measurements and analyses presented in the paper clearly indicate that the entire 6 GHz band is crucial for dense deployments, especially SP use, in order to meet growing demands for connectivity in crowded outdoor venues such as stadiums: our 6 GHz capable measurement phones were always connected to 6 GHz during game-day since only 14\% of the clients had this capability and hence were likely on 5 GHz. We also demonstrated that even with the SP APs transmitting at a median level of 27 dBm, 9 dB less than the allowed 36 dBm, and 6 dB more than the LPI level of 21 dBm, there was significant impact on indoor LPIs during game-day in areas where there is less isolation between the two deployments, e.g. outdoors and indoors near windows. The use of CSMA/CA in both outdoor and indoor deployments leads to a natural adaptation to the deployment environment, which may not be the case, for example, in the hybrid sharing proposed by Ofcom~\cite{ofcom20236ghz} where the outdoor deployment is 5G with much higher power than LPI and does not use CSMA/CA: our measurements indicate that frequencies used with high power outdoors may be unavailable for LPI use indoors, especially near windows. Hybrid sharing in 6 GHz should consider the use of 5G NR-U instead, which does have a similar protocol as CSMA/CA~\cite{naik2020next}. Finally, while the AFC does not consider aggregate interference while calculating allowed power, our spectrum analyzer measurements indicate a rise of about 10 dB in aggregate signal strength during game-day: it should be noted that the deployment studied was extremely dense with approximately 100 BSSIDs on each channel and this aggregate interference will be much lower in most cases. However, there may be a need for the AFC to evaluate allowed transmit power in dense deployments differently.

We will continue our measurement and analysis work to monitor the evolution of 6 GHz Wi-Fi deployment. The work reported here focused on analyzing spectrum coexistence in different scenarios: our future work will focus on similarly detailed comparisons of throughput (uplink and downlink) and latency performance of 5 GHz and 6 GHz Wi-Fi in dense deployments.

\section*{\centering{Acknowledgements}}

This work is supported by a research grant from the Department of Energy (DOE) Office of the Cybersecurity, Energy Security, and Emergency Response (CESER), in collaboration with Idaho National Lab (INL), and NSF grants CNS-2229387, CNS-2346413 and AST-2132700.

%This research is supported in part by the National Science Foundation under grant number CNS-2128489, 2132700, 2220286, 2220292, 2226437, and 2229387.

\bibliographystyle{IEEEtran}
\bibliography{main}

\end{document}